\documentclass[cernpreprint, backref=false, texlive=2020, UKenglish, dvipsnames, block=none, texmf]{atlasdoc}

\usepackage{atlaspackage}
\PreprintIdNumber{CERN-EP-2021-157}
 
\usepackage{atlasbiblatex}
 
\usepackage{atlasphysics}
 
\graphicspath{{logos/}{figures/}}
 
\usepackage{ANA-HDBS-2021-03-PAPER-defs}
\usepackage{atlassnippets}
\usepackage{atlasmisc}
\usepackage{atlasjetetmiss}
 
\AtlasTitle{Search for Higgs boson decays into a pair of pseudoscalar particles in the $bb\mu\mu$ final state with the ATLAS detector in $pp$ collisions at $\sqrt{s}=13$~TeV}
 
\AtlasVersion{2.0}
\AtlasAbstract{
This paper presents a search for decays of the Higgs boson with a  mass of 125~GeV into a pair of new pseudoscalar particles, $H\rightarrow aa$, where one $a$-boson decays into a $b$-quark pair and the other into a muon pair.
The search uses 139~fb$^{-1}$ of proton--proton collision data at a center-of-mass energy of $\sqrt{s}=13$~TeV recorded between 2015 and 2018 by the ATLAS experiment at the LHC. A narrow dimuon resonance is searched for in the invariant mass spectrum between 16~GeV and 62~GeV.
The largest excess of events above the Standard Model backgrounds is observed at a dimuon invariant mass of 52~GeV and corresponds to a local (global)  significance of $3.3 \sigma$ ($1.7 \sigma$).
Upper limits at  95\% confidence level are placed on the
branching ratio of the Higgs boson to the $bb\mu\mu$ final state, $ \mathcal{B}(H\rightarrow aa\rightarrow bb\mu\mu)$, and are in the range $\text{(0.2--4.0)} \times10^{-4}$, depending on the signal mass hypothesis.
 
}
 
\AtlasJournal{PRD}
\author{The ATLAS Collaboration}
 
\AtlasRefCode{HDBS-2021-03}

\AtlasJournalRef{Phys. Rev. D 105 (2022) 012006}
\AtlasDOI{10.1103/PhysRevD.105.012006}
\hypersetup{pdftitle={ATLAS document},pdfauthor={The ATLAS Collaboration}}
 
\begin{document}
 
\maketitle

\section{Introduction}

Light (pseudo) scalars that couple to the 125~\GeV\ Higgs boson~\cite{HIGG-2012-27,CMS-HIG-12-028} appear in many well-motivated extensions of the Standard Model (SM)~\cite{Dobrescu:2000yn,Ellwanger:2003jt,Dermisek:2005ar,Chang:2008cw,Morrissey:2008gm,Belanger:2013kya}.
These include models addressing the baryon asymmetry of the universe~\cite{Profumo:2007wc,Blinov:2015sna}, offering a solution to the naturalness problem~\cite{Craig:2015pha,Curtin:2015fna}, or providing insights into the nature of dark matter~\cite{Silveira:1985rk,Pospelov:2007mp,Draper:2010ew,Ipek:2014gua,Martin:2014sxa,Boehm:2014hva,GammaRayCascade}. Light bosons produced in Higgs boson decays could also be mediators to dark sectors that do not couple to the SM otherwise~\cite{Curtin:2013fra,TaniaTim1,TaniaTim2,TaniaTim3,ALPS}. Furthermore, pseudoscalar mediators appear in models, such as those described in Ref.~\cite{Liu_2019}, that were proposed to explain the anomalous muon magnetic moment~\cite{mug2_Fermi}.
A combination of ATLAS measurements of the Higgs boson production cross-sections and branching ratios
constrains the branching ratios  into invisible and undetected states to be  $\mathcal{B}(H\rightarrow\text{inv})<30\%$ and $\mathcal{B}(H\rightarrow\text{undetected})<21\%$, respectively,
whereas the overall branching fraction of the Higgs boson into beyond-the-SM (BSM) states is determined to be less than 47\% at 95\% confidence level (CL)~\cite{HIGG-2018-57}.
Combined measurements of Higgs boson couplings performed by the  CMS Collaboration set upper limits  of $\mathcal{B}(H\rightarrow\text{inv})<22\%$ and $\mathcal{B}(H\rightarrow\text{undetected})<38\%$ at 95\% CL~\cite{CMS-HIG-17-031}.
This motivates searches for light states in the Higgs boson decays that probe this potentially large $\mathcal{B}(H\rightarrow\text{BSM})$.

This paper presents a search for decays of the 125~\GeV\ Higgs boson into two pseudoscalars, denoted by $a$, in proton--proton ($pp$) collisions at the LHC~\cite{Evans:2008zzb}.
The search is performed in events where one $a$-boson decays into two $b$-quarks and the other into two muons, $H\rightarrow aa \rightarrow b\bar{b}\mu^+\mu^-$.\footnote{Denoted by $\haabbmm$ from now on for the rest of the paper.} The $a$-bosons are assumed to have  a decay width that is narrow compared to the detector resolution.
As pseudoscalar couplings are generally proportional to  mass,
which is for example the case in two-Higgs-doublet models~\cite{Curtin:2013fra,Branco:2011iw},
the $bb\mu\mu$ final state provides a good balance between a high branching ratio from the $a\to bb$  decay and a clean, high mass-resolution, dimuon resonance signature that is easy to trigger on from the $a\to\mu\mu$  decay.
In scenarios with enhanced lepton couplings,
the $a\to\mu\mu$ branching ratio can also be relatively large, resulting in $\mathcal{B}(H\rightarrow aa\rightarrow bb\mu\mu)/ \mathcal{B}(H\rightarrow aa )$ of  up to $ 0.16\%$~\cite{Curtin:2014pda}.
 
Light resonances in Higgs boson decays have been searched for by ATLAS and CMS in many different channels, i.e.\ in the  final states involving $4\mu$~\cite{CMS-HIG-18-003,EXOT-2016-22}, $2\mu2\tau$ or $4\tau$~\cite{CMS-HIG-18-024,CMS-HIG-18-006,CMS-HIG-17-029,CMS-HIG-16-015,HIGG-2014-02}, $2b2\tau$~\cite{CMS-HIG-17-024}, $4b$~\cite{HIGG-2017-05,HDBS-2018-47}, $4\gamma$~\cite{EXOT-2013-24} and $2\gamma+2$-jets~\cite{HIGG-2017-09}.
A search for a dimuon resonance produced in association with b-jets has been performed by CMS~\cite{CMS:2018lce} and a light resonance decaying to two muons has been searched for by LHCb~\cite{LHCb:2020ysn}.
CMS has performed a search for \haabbmm in 35.9~$\ifb$ of $pp$ collision data at a center-of-mass energy of $\sqrt{s}=13$~\TeV\ that sets upper limits on $\mathcal{B}(H \rightarrow aa \rightarrow bb\mu\mu)$ of $\text{(1--7)}\times 10^{-4}$ for $a$-boson masses ($m_a$) in the range $20 \leq m_a \leq 62.5$~\GeV~\cite{CMS-HIG-18-011}.
The ATLAS search  based on 36~$\ifb$ of Run~2 data~\cite{bbmm36ifb} sets upper limits on $\mathcal{B}(H \rightarrow aa \rightarrow bb\mu\mu)$ between $1.2\times10^{-4}$ and $8.4\times10^{-4}$ for $a$-boson masses  in the range $20 \leq m_a \leq 60$~\GeV.
In this paper, the full Run~2 dataset corresponding to an integrated luminosity of 139~$\ifb$ is used and the search is extended down to $m_a=16$~\GeV\ and up to $m_a=62$~\GeV.
Additionally,   boosted decision tree (BDT) techniques are used to improve the separation of  the signal from the SM backgrounds, increasing the analysis sensitivity, especially for higher $m_a$.

 
\section{ATLAS detector}

The ATLAS experiment~\cite{PERF-2007-01,PIX-2018-001} is a multipurpose particle detector
with a forward--backward symmetric cylindrical geometry and a nearly \(4\pi\) coverage in
solid angle.\footnote{ATLAS uses a right-handed coordinate system with its origin at the nominal interaction point (IP)
in the center of the detector and the \(z\)-axis along the beam pipe.
The \(x\)-axis points from the IP to the center of the LHC ring,
and the \(y\)-axis points upwards.
Cylindrical coordinates \((r,\phi)\) are used in the transverse plane,
\(\phi\) being the azimuthal angle around the \(z\)-axis.
The pseudorapidity is defined in terms of the polar angle \(\theta\) as \(\eta = -\ln \tan(\theta/2)\).
Angular distance is measured in units of \(\Delta R \equiv \sqrt{(\Delta\eta)^{2} + (\Delta\phi)^{2}}\).} It consists of an inner  detector (ID) surrounded by a thin superconducting solenoid
providing a \SI{2}{\tesla} axial magnetic field, electromagnetic (EM) and hadron calorimeters, and a muon spectrometer (MS).
The inner tracking detector covers the pseudorapidity range \(|\eta| < 2.5\).
It consists of silicon pixel, silicon microstrip, and transition radiation tracking detectors.
Lead/liquid-argon (LAr)  sampling calorimeters provide electromagnetic energy measurements
with high granularity.
A steel/scintillator-tile hadron calorimeter covers the central pseudorapidity range $|\eta| < 1.7$.
The endcap and forward regions are instrumented with LAr calorimeters
for both the EM and hadronic energy measurements up to \(|\eta| = 4.9\).
The MS surrounds the calorimeters and is based on
three large superconducting air-core toroidal magnets with eight coils each.
The field integral of the toroids ranges between \num{2.0}~Tm and \SI{6.0}Tm
across most of the detector.
The MS includes a system of precision tracking chambers and fast detectors for triggering.
A two-level trigger system is used to select events.
The first-level trigger is implemented in hardware and uses a subset of the detector information
to accept events at a rate below \SI{100}{\kHz}~\cite{TRIG-2016-01}.
This is followed by a software-based trigger that
reduces the accepted event rate to \SI{1}{\kHz} on average.
An extensive software suite~\cite{ATL-SOFT-PUB-2020-001} is used in the reconstruction and analysis of real and simulated data, in detector operations, and in the trigger and data acquisition systems of the experiment.

\section{Dataset and simulated events}
 
The data used in this analysis were collected in Run~2 of the LHC during the 2015--2018 data-taking period with $pp$ collisions at a center-of-mass energy of $\sqrt{s} = 13$~\TeV. The dataset corresponds to an integrated luminosity of 139~$\text{fb}^{-1}$.
The lowest-threshold unprescaled single-muon and dimuon triggers are used to select the events~\cite{TRIG-2018-01}. Single-muon triggers require the transverse momentum (\pT) of the muon  to be above 20~\GeV\ or 26~\GeV, depending on the data-taking period, while the dimuon trigger requires both muons to have a \pT above 14~\GeV.
 
Simulated events are used in the estimation of the SM backgrounds. \textsc{Sherpa}~2.2.1~\cite{Gleisberg:2008ta,Bothmann:2019yzt} was used as the baseline generator for the Drell--Yan (DY)\,+\,jets, $W(\rightarrow \ell \nu)$+jets, diboson and triboson backgrounds. It is a multiparton matrix element and parton shower (PS) generator including hadronization~\cite{Schumann:2007mg,Hoeche:2011fd, Hoeche:2012yf, Catani:2001cc,Hoeche:2009rj}, with the \textsc{NNPDF3.0} parton distribution function (PDF) set at next-to-next-to-leading-order (NNLO) accuracy~\cite{Ball:2014uwa}.
The DY+jets and multiboson samples were generated with a minimum dilepton mass of 10~\GeV\ and 4~\GeV, respectively.
The \ttbar and single-top-quark samples were generated with \textsc{Powheg-Box~v2}~\cite{Frixione:2007vw,Nason:2004rx,Alioli:2010xd,Frixione:2007nw,Re:2010bp} using the \textsc{NNPDF3.0nlo} PDF in matrix element interfaced to \textsc{Pythia~8.230}~\cite{Sjostrand:2014zea} for the PS.
For the underlying-event description a set of tuned parameters called the \textsc{A14} tune~\cite{A14_article} was used, along with the \textsc{NNPDF2.3lo} PDF~\cite{Ball:2012cx}.
The \ttbar + vector-boson processes ($\ttbar{+}V$) were generated with \textsc{MadGraph5}\_aMC@NLO~2.3.3~\cite{Alwall:2014hca} interfaced to \textsc{Pythia}~8.210 for the PS. The underlying-event tune was the same as for the \ttbar sample.
\textsc{EvtGen}~\cite{Lange:2001uf} was used for the properties of the bottom and charm hadron decays in all simulated samples, except those simulated with \textsc{Sherpa}.

Higgs boson production through gluon--gluon fusion (ggF) was generated using the NNLOPS program~\cite{Hamilton:2012np,Hamilton:2015nsa} with \textsc{Powheg-Box~v2}~\cite{Frixione:2007vw,Alioli:2010xd, Campbell:2012am, Hamilton:2012rf}. The vector-boson fusion (VBF) processes were generated with \textsc{Powheg-Box~v2} at NLO accuracy~\cite{Nason:2009ai}. The Higgs boson mass was set to 125~\GeV. For both the ggF and VBF production processes, \textsc{Powheg-Box} was interfaced with \textsc{Pythia}~8.212 using  the AZNLO tune~\cite{STDM-2012-23}
for the simulation of the $H\rightarrow aa\rightarrow bb\mu\mu$ decays, where the $a$-boson is a pseudoscalar, as well
as for parton showering, hadronization and the underlying event.
The ggF Higgs boson production rate is normalized to the total cross-section predicted at
next-to-next-to-next-to-leading-order accuracy in QCD with NLO electroweak corrections applied~\cite{deFlorian:2016spz,Anastasiou:2015ema,Anastasiou:2016cez,Actis:2008ug,Anastasiou:2008tj} and amounts to 48.58~pb.
The VBF production rate is normalized to an approximate NNLO cross-section with the NLO electroweak corrections applied~\cite{Heinemeyer:2013tqa,Ciccolini2007,Ciccolini2008,BolzoniVBF}, which amounts to 3.8~pb.
The contribution from the associated production of a Higgs boson and a vector boson (VH) is calculated to be 3.5\% of the total  ggF+VBF cross-section and is accounted for by scaling the simulated ggF and VBF samples.
The contribution from Higgs boson production in association with a pair of top quarks is found to be negligible (below the percent level) and is neglected in the analysis.
Thirteen mass points were simulated for the ggF and VBF production modes, with the $a$-boson mass in the range $m_a = \text{16--62}$~\GeV.\footnote{More specifically, the simulated mass points are at $m_a = 16$, 18, 20, 25, 30, 35, 40, 45, 50, 52, 55, 60 and 62~\GeV.}
Below $m_a = 16$~\GeV\ the $b$-quarks coming from the decays of the $a$-boson tend to be so collimated due to its boost that they cannot be reconstructed as two separate $b$-jets (with a radius parameter of $R=0.4$). Another effect is that in the highly asymmetric decays of low-mass $a$-bosons, the subleading $b$-jet falls below the jet reconstruction threshold of 20~\GeV~\cite{PERF-2015-09}.
As a result, the signal acceptance falls below 0.2\% and the analysis loses sensitivity.

The effects of additional interactions in the same and neighboring beam-bunch crossings (pileup) were modeled for all simulated events by overlaying additional $pp$ collisions generated with \textsc{Pythia~8.186} using the \textsc{NNPDF2.3lo} PDF set and the A3 tune~\cite{Sjostrand:2007gs}. Simulated event samples are weighted to reproduce the distribution of the number of pileup interactions observed in the data. All the generated background and signal samples are processed through the  ATLAS detector simulation~\cite{SOFT-2010-01} based on $\GEANT4$~\cite{Agostinelli:2002hh} and reconstructed using the same software as for the data.

\section{Event reconstruction and selection}

Muons are reconstructed by combining track information from the MS with tracks found in the ID \cite{MUON-2018-0333}. They also have to satisfy $\pt>5$~\GeV\ and $|\eta|<2.7$ (for $|\eta|>2.5$, only tracking information from the MS is used), and pass the $LowPt$ working point identification requirement defined in Ref.~\cite{MUON-2018-0333}.
Muon tracks must have a longitudinal impact parameter $z_0$ satisfying $|z_{0}\sin\theta|<0.5$~mm and a transverse impact parameter significance $|d_{0}|$/$\sigma_{d_{0}}<3$ relative to the primary interaction vertex, chosen as the reconstructed vertex with the highest sum of the $\pT^2$ of its associated tracks. Furthermore, muons are required to be isolated from the surrounding detector activity by requiring that the scalar sum of the \pt of additional inner detector tracks  and the sum of the  transverse momentum $E_\mathrm{T}$ of calorimeter energy deposits within a cone of size $\Delta R = 0.2$ around a muon be less than 15\% and 30\% of the muon \pt, respectively.

Jets are reconstructed using the anti-$k_t$ algorithm~\cite{Cacciari:2008gp} implemented in the \textsc{FastJet}\xspace package~\cite{Fastjet} with a radius parameter of $R = 0.4$. The inputs to the jet clustering are built by combining the information from both the calorimeters and the ID using a particle-flow algorithm~\cite{PERF-2015-09,JETM-2018-05}. Jets with $\pT < 60$~\GeV\ originating from pileup are suppressed with the jet-vertex-tagger (JVT)~\cite{ATLAS-CONF-2014-018}, a multivariate algorithm combining track-based variables. Selected jets are required to have $\pt > 20$~\GeV\ and $|\eta| <2.5$.
An algorithm (MV2c10) relying on multivariate techniques, taking as input the properties of displaced tracks and vertices reconstructed within a jet, is employed to identify (tag) jets containing $b$-hadrons~\cite{FTAG-2018-01}.
The MV2c10 tagger is used at 77\% $b$-jet identification efficiency, with an approximate misidentification probability of 25\% for jets arising from charm quarks, 6.3\% for hadronically decaying $\tau$-leptons, and 0.8\% for light-flavor jets as measured in simulated \ttbar events.
 
The missing transverse momentum (\met) is calculated as the magnitude of the negative vector sum of the transverse momenta of all the reconstructed and calibrated objects in the event, including a soft term that accounts for charged particles that are associated with the primary vertex, but not with any reconstructed object~\cite{PERF-2016-07,ATLAS-CONF-2018-023}.

The events selected for the analysis are required to have two muons of opposite charge, either with the leading and subleading muons satisfying $\pt^{\text{leading}}>27$~\GeV\ and $\pt^{\text{subleading}}>5$~\GeV, and the event being triggered by a single-muon trigger, or with both muons having $\pt >15$~\GeV, and the event being triggered by a dimuon trigger. The dimuon invariant mass, $\Mmm$, is required to be between 15~\GeV\ and 65~\GeV. Furthermore, the events must contain exactly two $b$-tagged jets with \pT above 20~\GeV.

A kinematic likelihood (KL) \cite{Erdmann:2013rxa} fit exploiting the equal invariant masses of the $bb$ and $\mu\mu$ systems in $\haa$ decays  is performed to improve the four-body invariant mass ($\Mbbmm$) resolution and  reduce the SM backgrounds. The same fit approach as considered in the previous ATLAS publication~\cite{bbmm36ifb} is used.
The dimuon invariant mass, $\Mmm$, is used to constrain the di-$b$-jet mass, as the former has a resolution approximately ten times better than the latter.
The $\Mmm$ resolution ranges between 0.4~\GeV\ at $m_a$ = 16~\GeV\ and 1.3~\GeV\ at $m_a = 62$~\GeV.
The fit maximizes the likelihood by shifting the $b$-jet energies within the resolution in order to satisfy the constraint $\Mmm \simeq \Mbb$.
The output of the fit is the logarithm of the maximum likelihood value, \LogL, which quantifies how well the event matches the $\Mmm = \Mbb$ hypothesis, characteristic of signal events.
The four-body invariant mass, recomputed after the KL fit, is denoted by $\KLMbbmm$ and is used for further event categorization.
 
Signal-like events are chosen by requiring that $110 <\KLMbbmm<140$~\GeV, and that $\LogL > -8$, which ensures that $\Mbb$ is compatible with $\Mmm$.
Finally, \met is required to be less than 60~\GeV\ to reduce the background from \ttbar events, which is one of the two major backgrounds and can contain large \met from neutrinos in top-quark decays. This  selection defines the ``inclusive'' signal region (SRincl) and is summarized in Table~\ref{selections}, along with the selection requirements for other analysis regions described later in the text.
 
A boosted decision tree (BDT) classifier implemented using the TMVA framework \cite{Speckmayer:2010zz} is employed to further reduce the SM backgrounds.
Its training is done in partially overlapping 8-\GeV-wide \Mmm windows centered at the $m_a$ values of each of the 12 generated signals,\footnote{One BDT was trained for each generated signal MC sample, except for $m_a = 52$~\GeV, as this sample was produced only at a later analysis stage.} in order to fully exploit their kinematic differences.
The background sample consists of \ttbar and DY+jets events, the two dominant backgrounds, combined in the proportions extracted from the background validation fit described in Section~\ref{sec:result}. The signal samples used for the training include ggF and VBF Higgs boson production samples combined according to their cross-sections.
The seven kinematic variables included in the training are:
\begin{itemize}
\item{$\Mbb$,}
\item{\LogL,}
\item{$\Delta R_{b_1b_2}$ (the angular distance between the two $b$-jets),}
\item{diff$\Delta R_bR_{\mu}=\Delta R_{b_1b_2}-\Delta R_{\mu_1\mu_2}$ (the difference between the angular separations between the two $b$-jets and the two muons),}
\item{$\Delta R_{bb\mu\mu}$ (the angular distance between the $bb$ and $\mu\mu$ systems),}
\item{$\overline{\Delta R_{b,\mu}} = \left[\Delta R_{b_1\mu_1}+\Delta R_{b_1\mu_2}+\Delta R_{b_2\mu_1}+\Delta R_{b_2\mu_2}\right]/4$ (the average angular distance  of all four combinations of a $b$-jet and a muon),}
\item{$\overline{m_{b\mu}}=\left[m_{b_1\mu_1}+m_{b_1\mu_2}+m_{b_2\mu_1}+m_{b_2\mu_2}\right]/4$ (the average mass of all four combinations of a $b$-jet and a muon).}
\end{itemize}
The distributions of these variables for the background and three representative signal masses are shown in Figure~\ref{BDTinputs}.
 
The $\Mbb$ variable helps separate the low-mass signal from the backgrounds, as $\Mbb$ peaks around 60~\GeV\ for the \ttbar and DY processes. The \LogL~peaks at higher values as the signal mass becomes smaller.
 
Due to a higher boost of a lighter $a$-boson, its decay products are collimated, resulting in $\Delta R_{b_1b_2}$ and  $\Delta R_{\mu_1\mu_2}$ being much smaller than for a signal from a heavier $a$-boson or for background processes.
As a consequence, diff$\Delta R_bR_{\mu}$ shows a narrow distribution centered around zero, while the background and a higher-mass signal exhibit a much broader diff$\Delta R_bR_{\mu}$ distribution.

The $\Delta R_{bb\mu\mu}$ variable helps enhance the sensitivity to higher signal masses. Heavier $a$-bosons are produced approximately at rest, resulting in the $\Delta R_{bb\mu\mu}$ distribution being relatively flat with a small peak at low values. As the signal mass decreases, the $\Delta R_{bb\mu\mu}$ distribution  transitions into a ``back-to-back'' topology, characteristic of both a low-mass signal and the background events.

Finally, the $\overline{\Delta R_{b,\mu}}$ and $\overline{m_{b,\mu}}$ variables provide another measure of how close the two $a$-bosons are in $\Delta R$.
In the ``back-to-back'' topology for lower signal masses, the muons are, on average, further away from the $b$-jets, while for heavier $a$-bosons produced approximately at rest, the average distance between the muons and the $b$-jets is smaller.
Consequently, $\overline{\Delta R_{b,\mu}}$ and $\overline{m_{b,\mu}}$ peak at high (low) values for low (high) signal masses, while the backgrounds peak somewhere between the two extreme signal topologies.
 
The output score of the BDT trained for a signal with mass $m_a$ is denoted by BDT$m_a$. The BDT$m_a$ distributions for $m_a$ = 20~\GeV, 40~\GeV\, and 60~\GeV\ are shown in Figure~\ref{BDToutputs}.
 
The final signal region (SR) bin for each signal mass is defined by imposing two requirements in addition to the SRincl selection: $m_a-X<\Mmm<m_a+X$ and  BDT$m_a>0.2$, where $X=1$~\GeV\ ($X=1.5$~\GeV) for $m_a\leq 45$~\GeV\ ($m_a>45$~\GeV). The widths of the SR bins and the BDT$m_a$ cut value are optimized to maximize the significance of signal over background events.
For masses at which no signal sample was generated, and, consequently, no BDT was trained, the BDT trained for the $m_a$ closest to the one being tested is used.
For example, when testing the  $m_a=32$~\GeV\ hypothesis, the requirement BDT$30>0.2$ is applied to select the events for the SR bin.
Signal yields for mass points where no signal sample was generated ($m_a=32$~\GeV\ in this example) are obtained by selecting events with BDT scores above 0.2 for the same BDT$m_a$ (BDT30 in this case) in all simulated mass points and interpolating using third-order splines.
To assess the uncertainty, the yields of the neighboring simulated mass points ($m_a=30$~\GeV\ and $m_a=35$~\GeV\ in this case) are interpolated using a linear function. The difference between the yields obtained using the splines and a linear function for the interpolation is assigned as a systematic uncertainty on the interpolated signal yield.
Using a BDT at a mass for which the training was not performed results in a negligible loss of significance relative to a BDT that was optimized for that mass point.

\begin{table}[h!]
\caption{Summary of the selection requirements for the  control (TCR and DYCR), validation (VR1 and VR2) and  inclusive signal (SRincl) regions in the analysis, as well as the final SR bins. The control and validation regions are defined in Section~\ref{sec:backgrounds}.}
\label{selections}
\begin{center}
\begin{tabular}{l|c|c|c|c|c}
\toprule
& TCR & DYCR & SRincl & VR1 & VR2 \\
\midrule
\midrule
\Mmm[\GeV] & \multicolumn{5}{c}{[15, 65]} \\
\midrule
\KLMbbmm[\GeV] & [110, 140] & [80, 110] or [140, 170]  & [110, 140]  & [170, 300] & [110, 140] \\
\midrule
\met[\GeV] &  $> 60$ & \multicolumn{4}{c}{$< 60$} \\
\midrule
\LogL & \multicolumn{4}{c|}{$> -8$} & [$-11$, $-8$] \\
\midrule
\midrule
SR bins & \multicolumn{5}{c}{  SRincl  \& BDT$m_a>0.2$ }  \\
& \multicolumn{5}{c}{ 2-\GeV-wide (3-\GeV-wide) \Mmm bins for $m_a\leq45$~\GeV\ ($m_a>45$~\GeV) }  \\
\bottomrule
\end{tabular}
\end{center}
\end{table}

The signal acceptance $\times$ efficiency varies between 0.3\% and 2.5\% for ggF Higgs boson production and between 0.2\% and 3.0\% for VBF production, where the lowest acceptance $\times$ efficiency is obtained for the lowest $m_a$, and grows as $m_a$ increases. The largest loss of acceptance occurs when requiring that there are two $b$-jets in the event, as one of the signal jets tends to fall below the reconstruction threshold of 20~\GeV. The fraction of signal events passing the two-$b$-jet requirement is less than 20\% for all mass points.

\begin{figure}
\centering
\includegraphics[width=0.45\hsize]{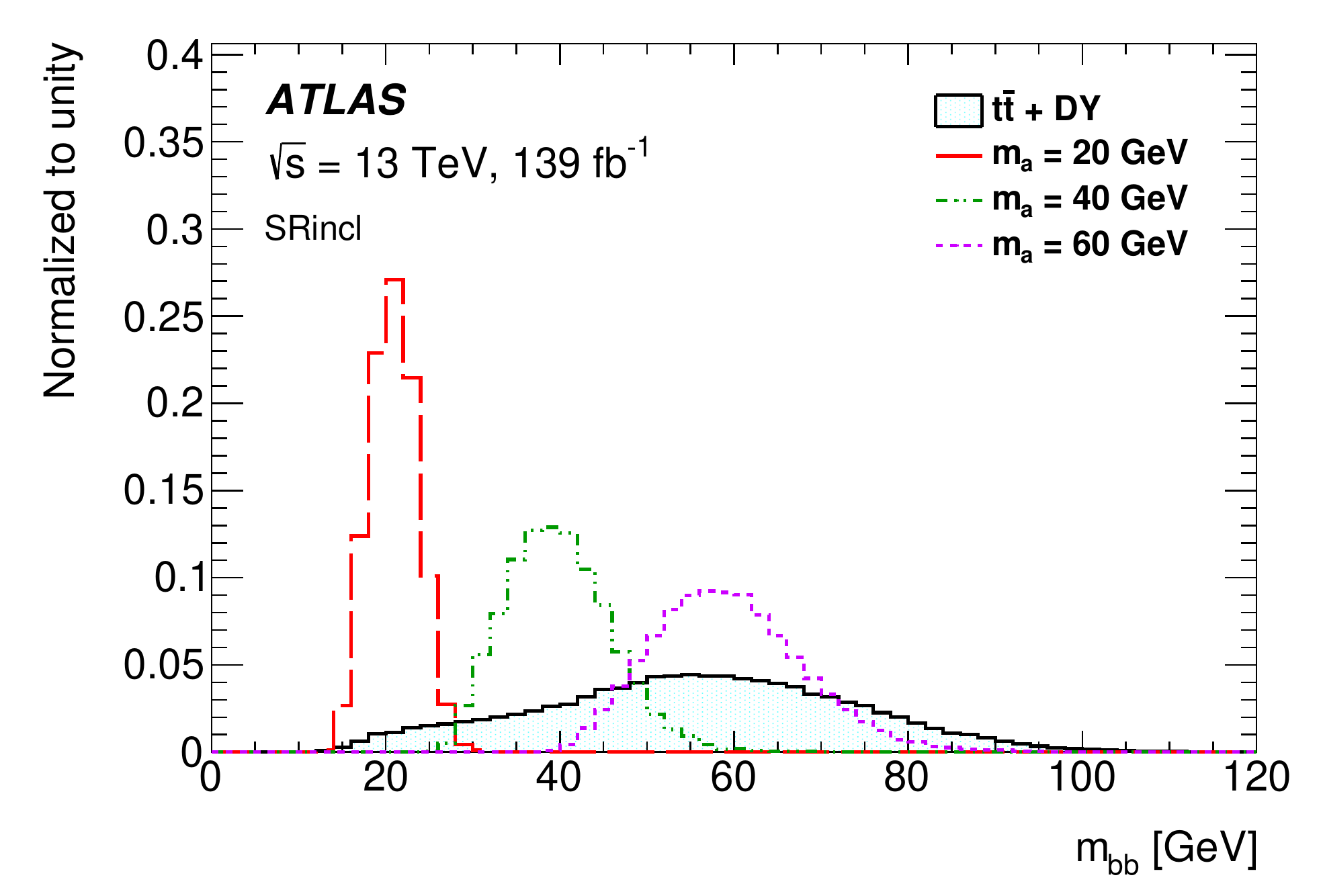}
\includegraphics[width=0.45\hsize]{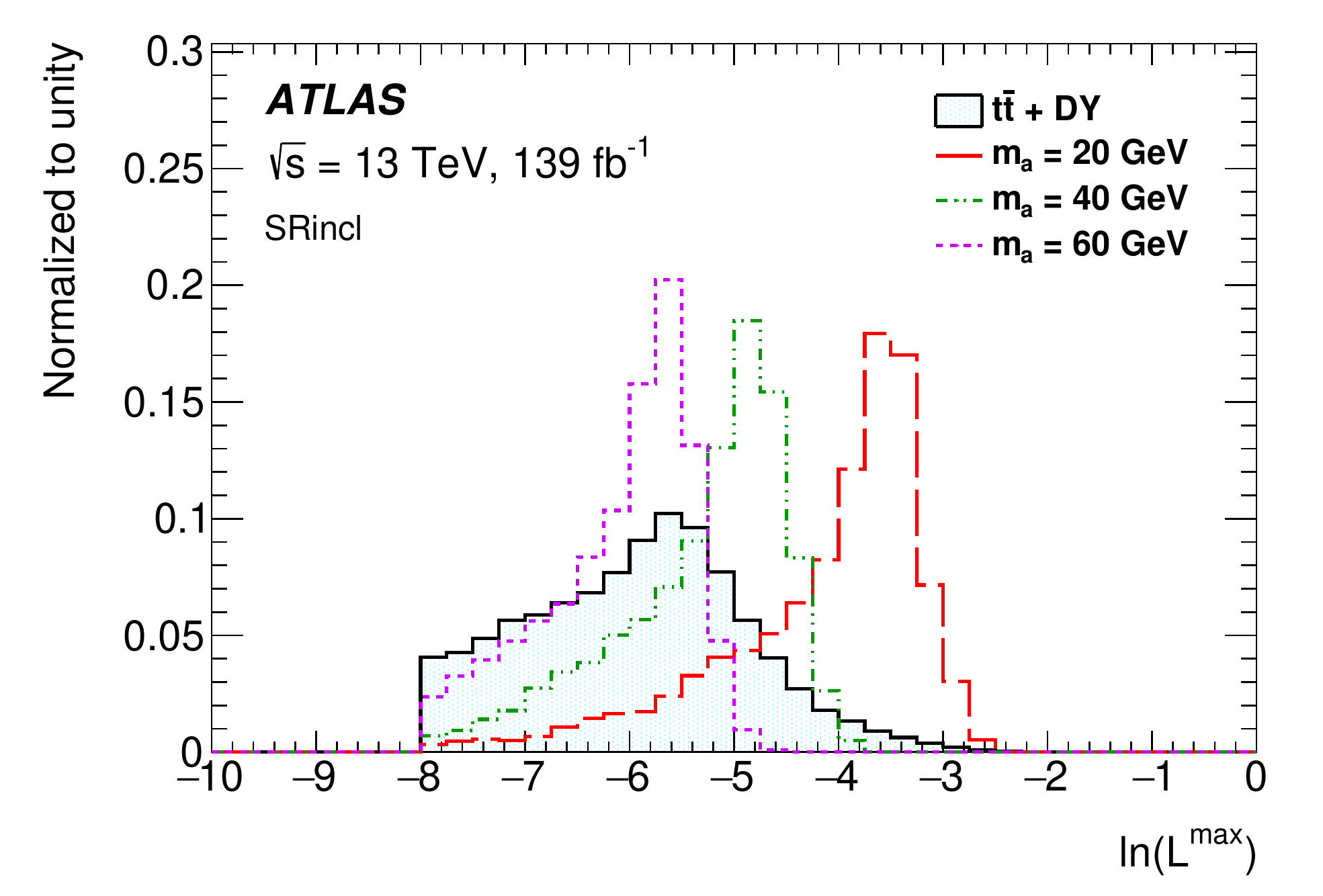}
\includegraphics[width=0.45\hsize]{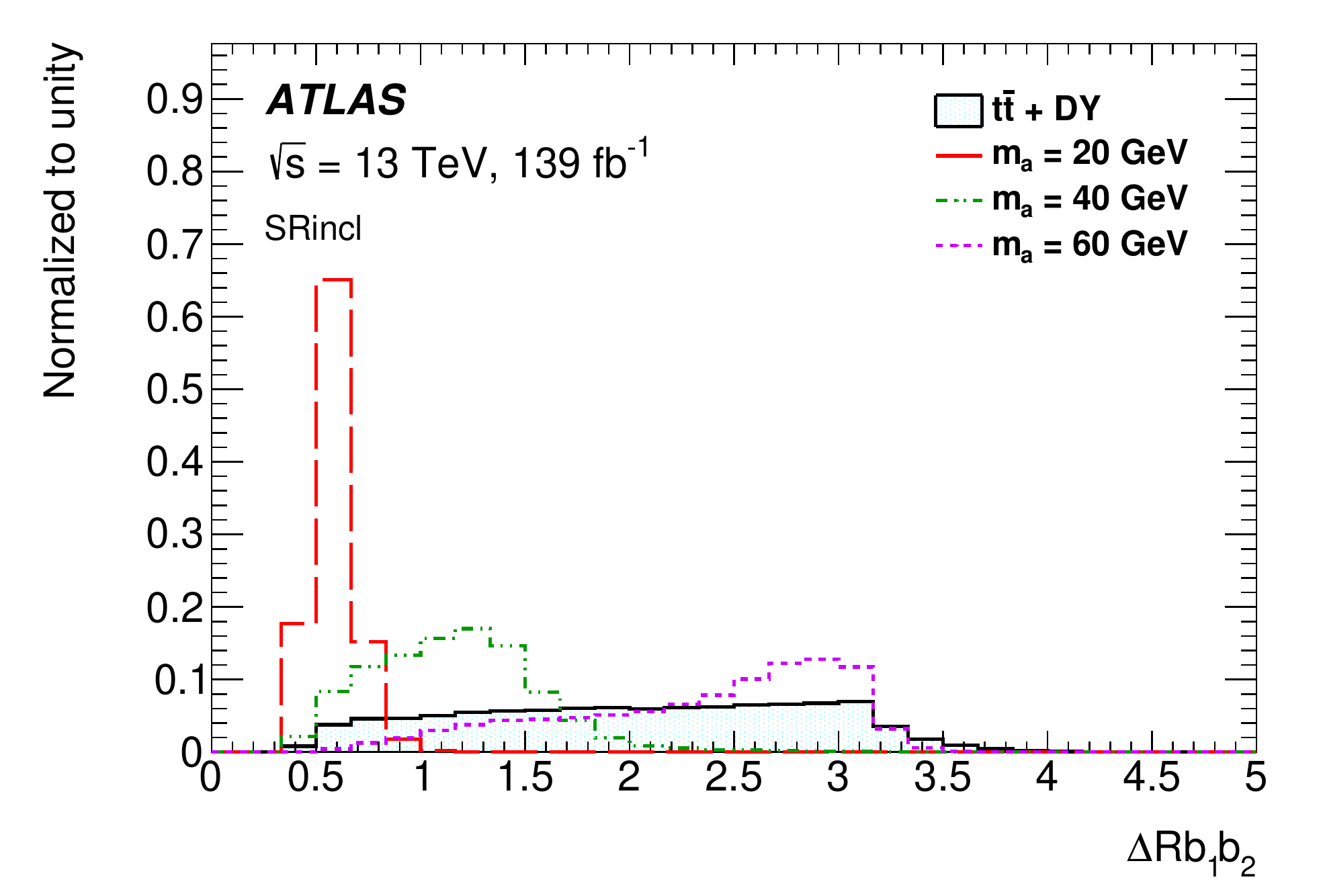}
\includegraphics[width=0.45\hsize]{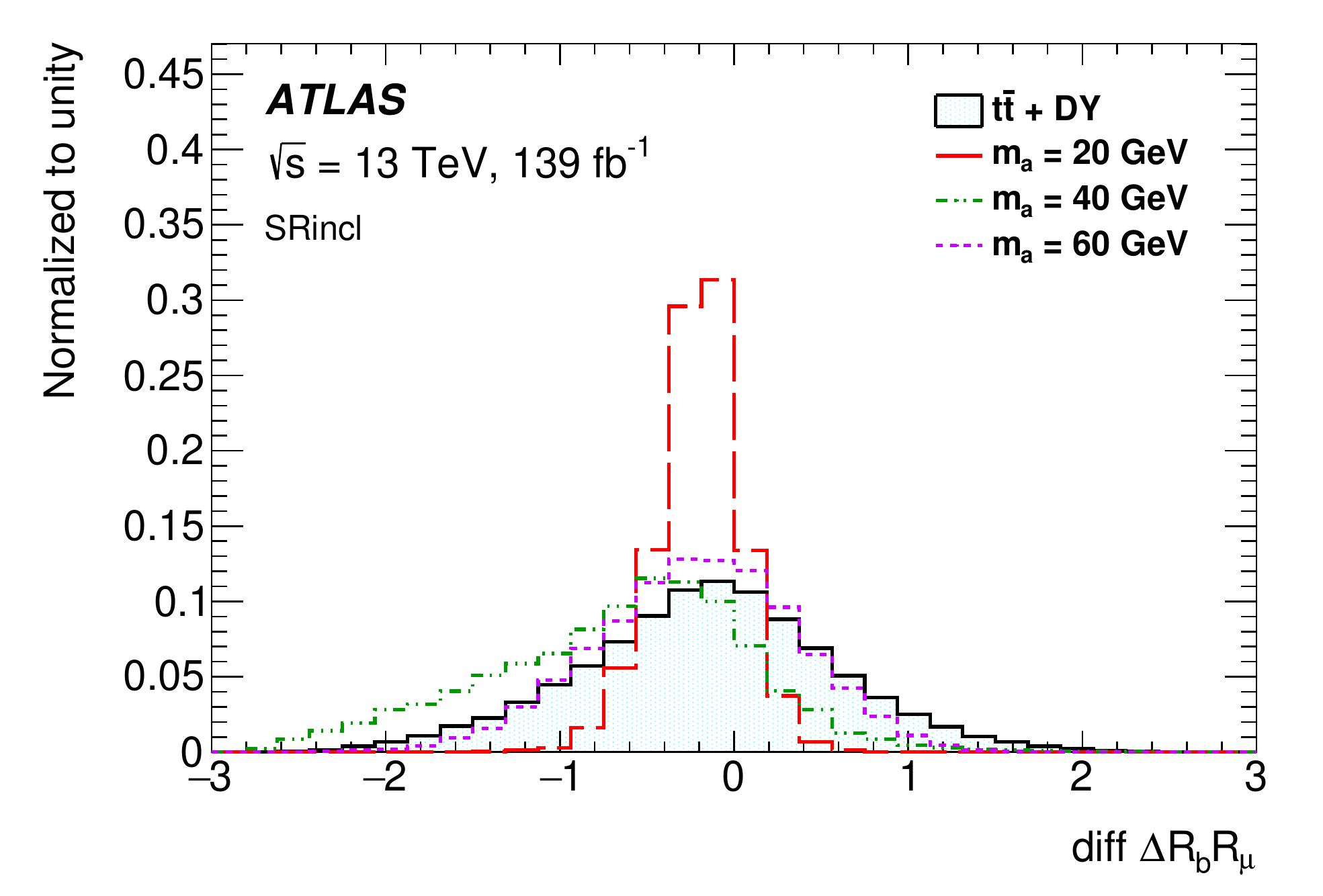}
\includegraphics[width=0.45\hsize]{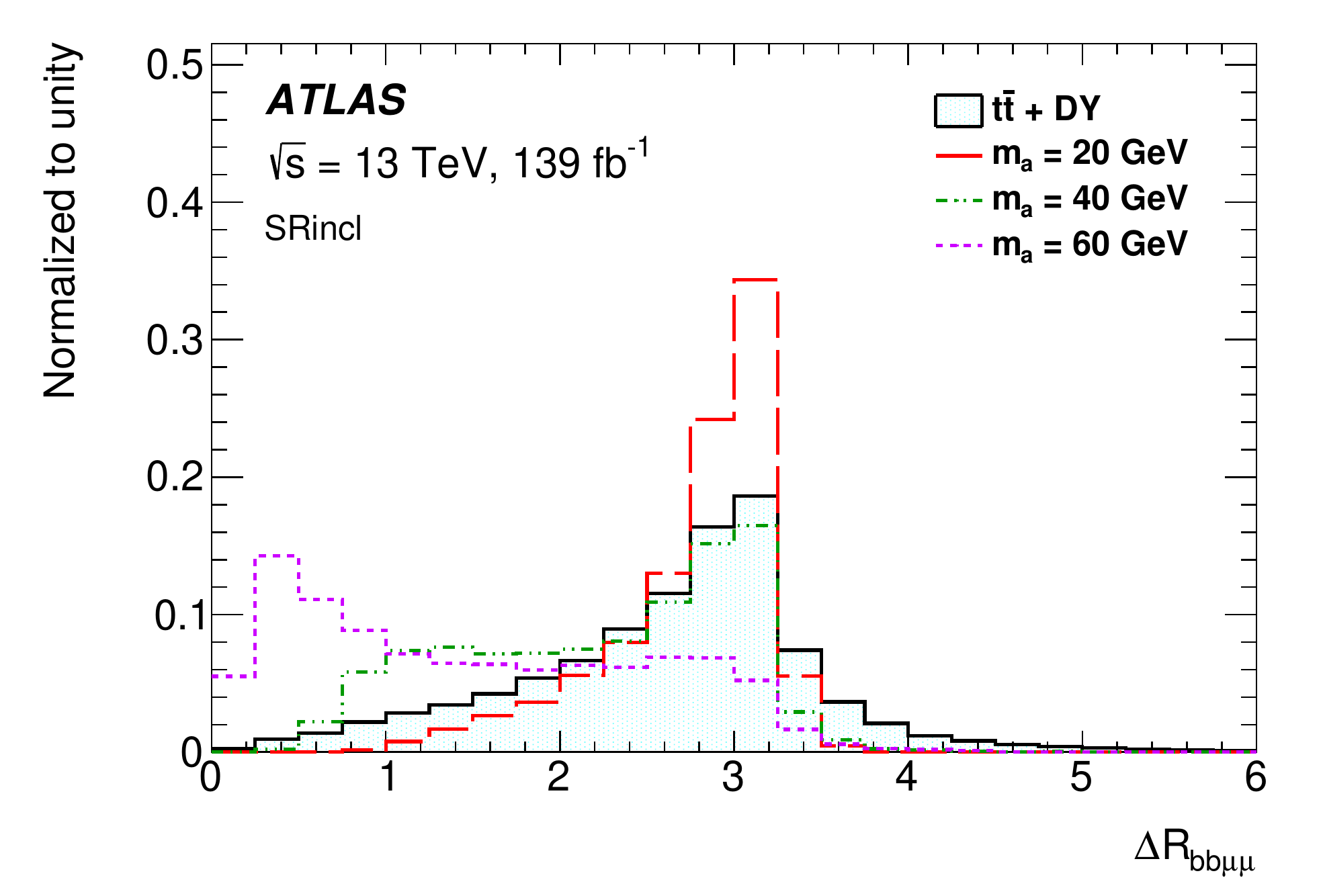}
\includegraphics[width=0.45\hsize]{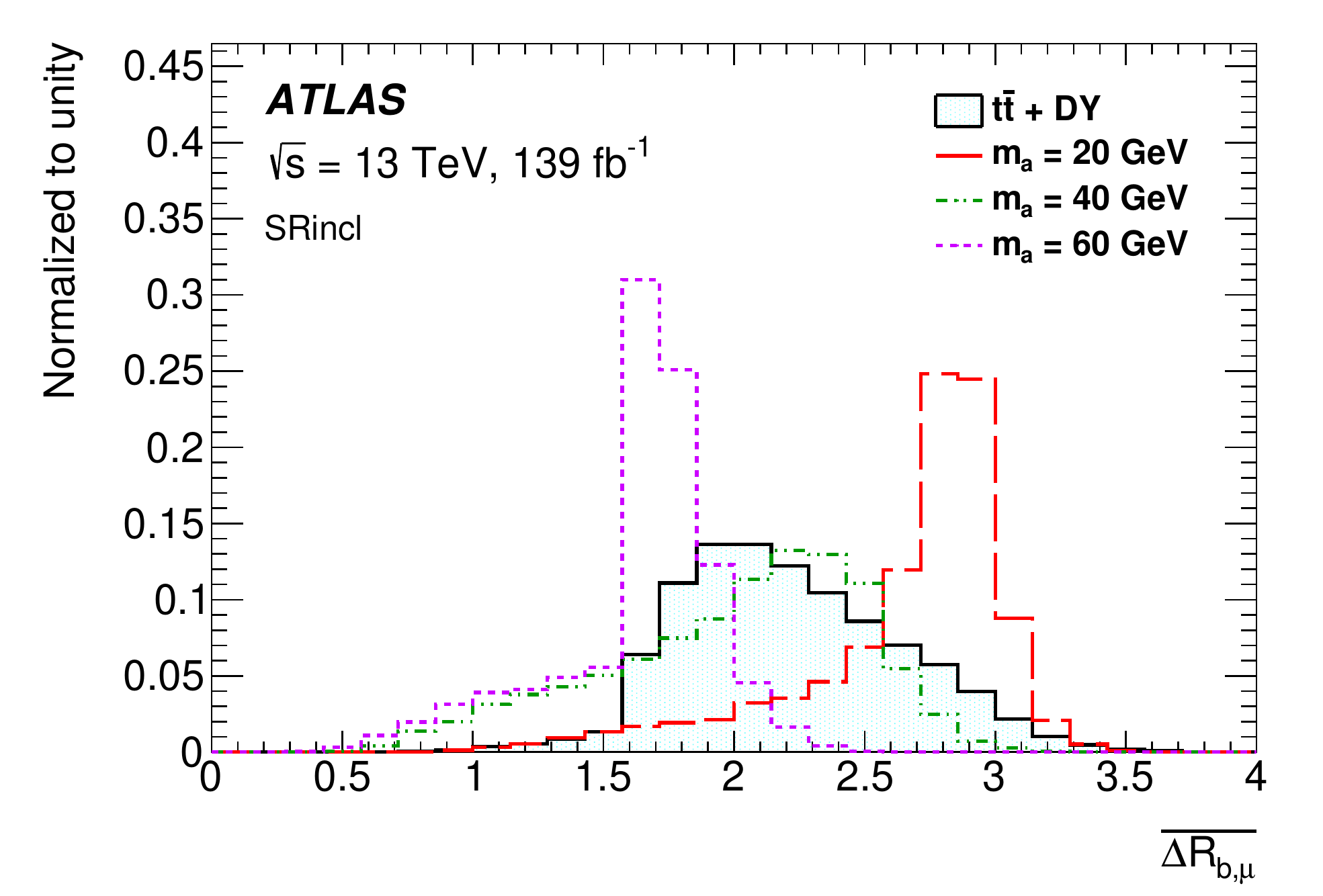}
\includegraphics[width=0.45\hsize]{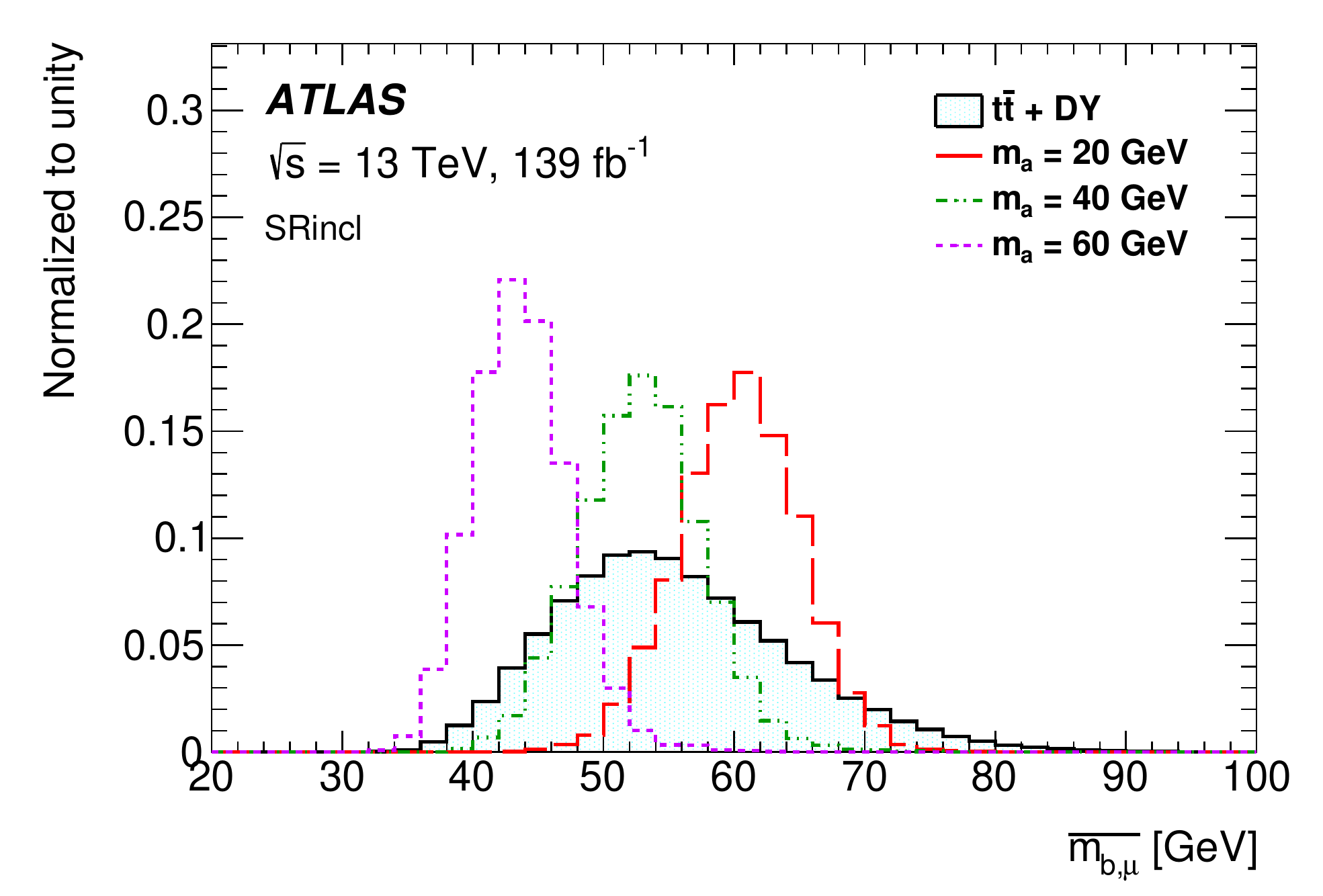}
\caption{Kinematic variables used as inputs to the BDT training. From top left to bottom right: $\Mbb$, \LogL, $\Delta R_{b_1b_2}$, diff$\Delta R_bR_{\mu}$, $\Delta R_{bb\mu\mu}$, $\overline{\Delta R_{b,\mu}}$, $\overline{m_{b,\mu}}$. The variables are plotted in SRincl.
All the distributions are normalized to unit area. The background histogram is the sum of the \ttbar and DY event templates, combined in the proportions extracted from the background validation fit described in Section~\ref{sec:result}.}
\label{BDTinputs}
\end{figure}
 
\begin{figure}
\centering
\includegraphics[width=0.45\hsize]{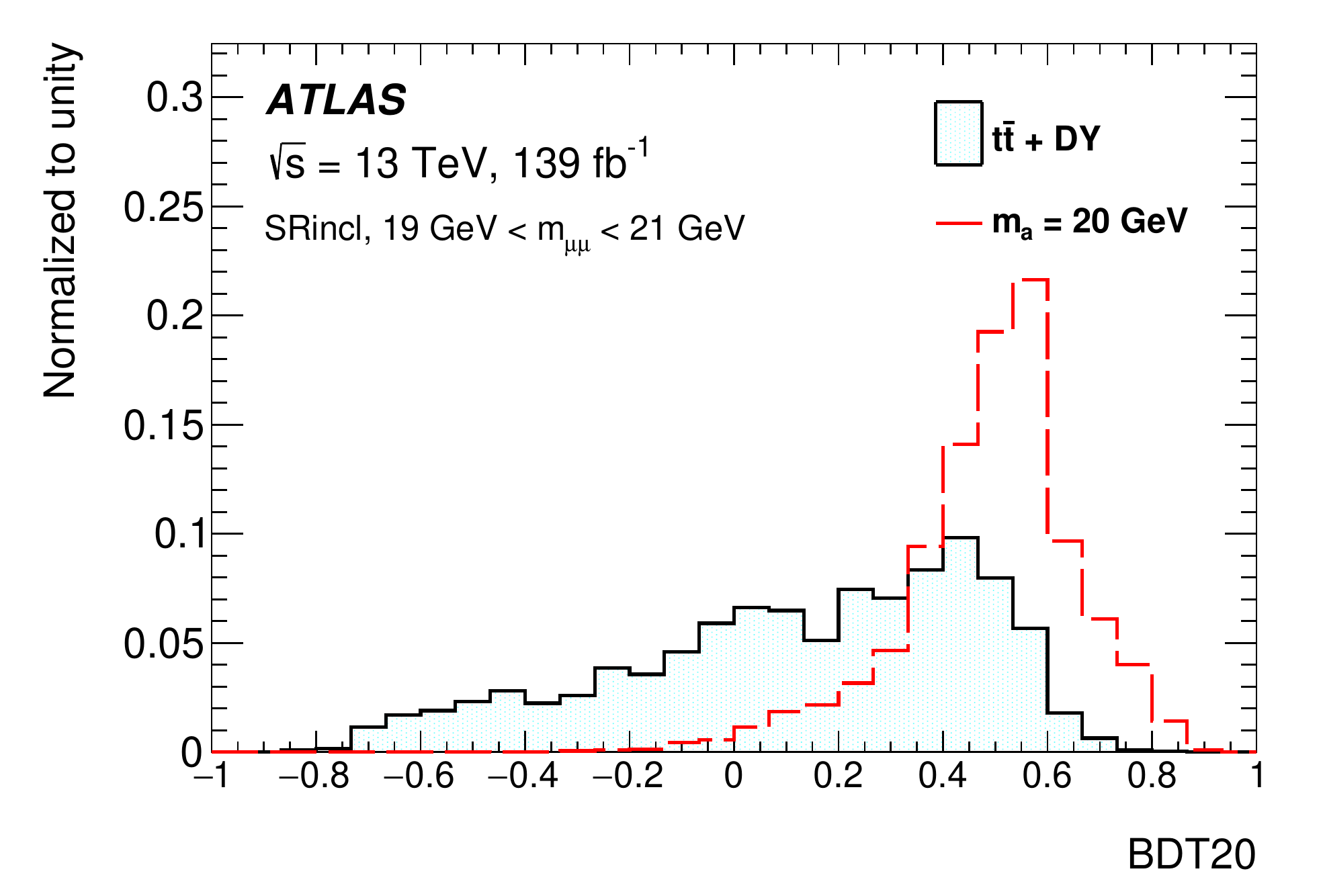}
\includegraphics[width=0.45\hsize]{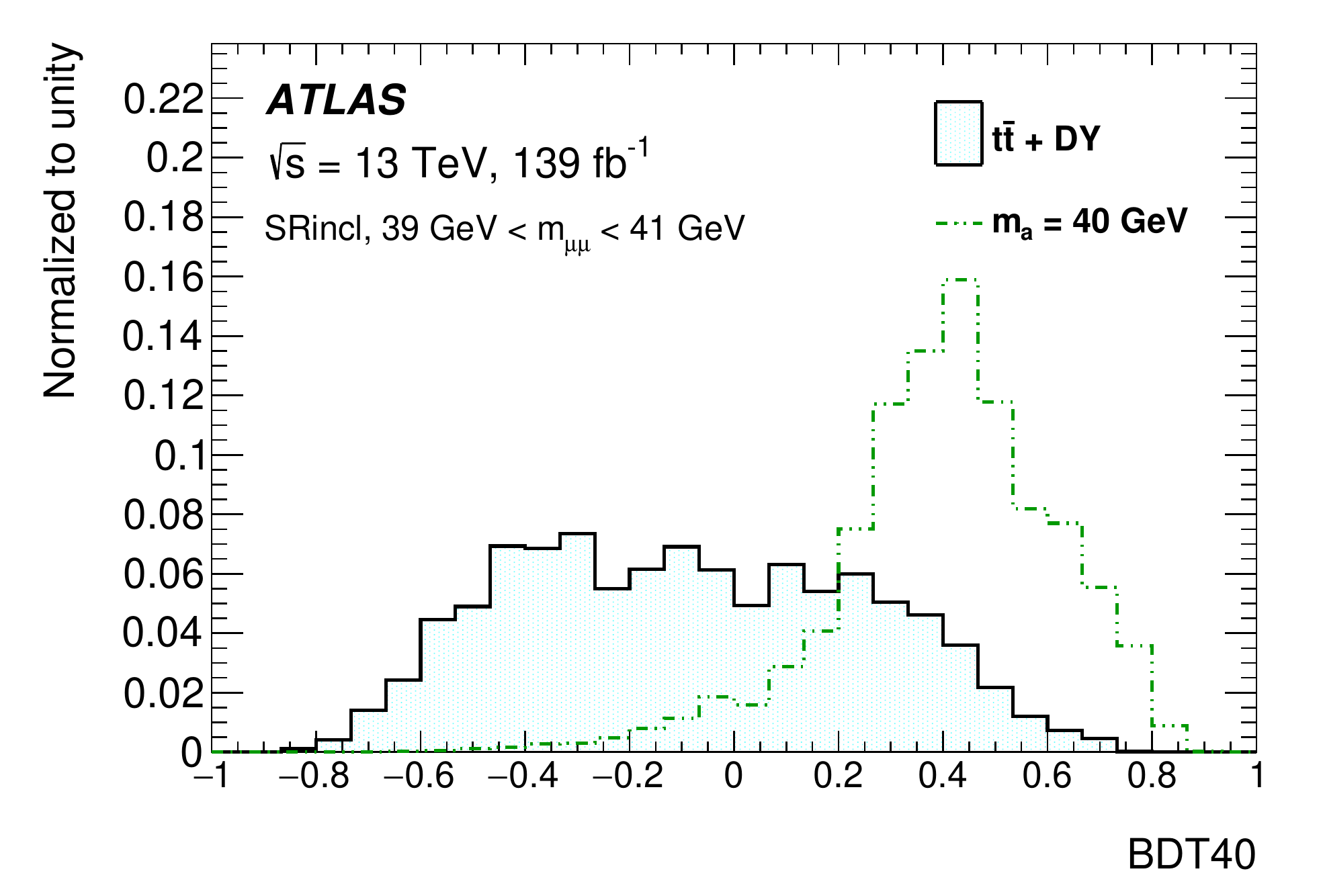}
\includegraphics[width=0.45\hsize]{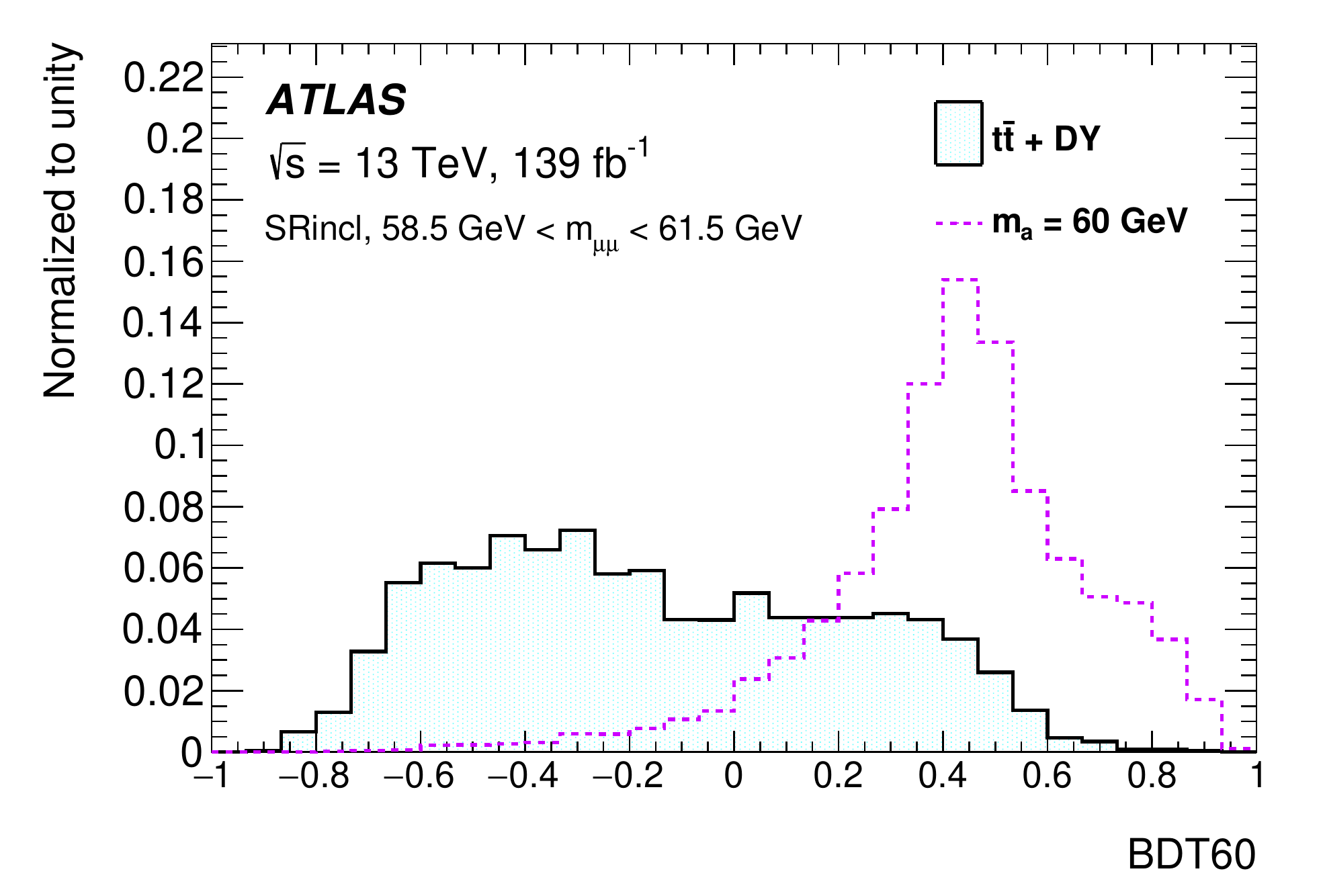}
\caption{Three BDT$m_a$ distributions, BDT20, BDT40, and BDT60, plotted in the \Mmm windows of SRincl, as indicated in the figures. The distributions are normalized to unit area. The background histogram is the sum of the \ttbar and DY event templates, combined in the proportions extracted from the background validation fit described in Section~\ref{sec:result}.}
\label{BDToutputs}
\end{figure}

\section{Background estimation}
\label{sec:backgrounds}
The dominant backgrounds in the analysis arise from the DY dimuon process in association with $b$-quarks and  pair production of top quarks ($\ttbar$) where each $W$ boson decays into a muon and a neutrino. These two backgrounds account for more than $96\%$ of background events in all analysis regions.
 
Two control regions are designed to constrain the  \ttbar and DY backgrounds. They are chosen so that they have negligible signal contamination, are kinematically as close as possible to SRincl, and maximize the contribution of one of the respective background processes.
A top-quark control region (TCR) is defined by inverting the $\met$ selection criterion in SRincl to $\met>60$~\GeV. This results in an event sample approximately 93\% pure in \ttbar events. The DY control region (DYCR) is defined in the 30 \GeV-wide \KLMbbmm sidebands of SRincl, i.e.\ by requiring $80 <\KLMbbmm<110$~\GeV\ or $140 <\KLMbbmm<170$~\GeV. Approximately 50\% of the events in DYCR originate from the DY process, whereas the rest mostly come from \ttbar production.
Two validation regions (VR1 and VR2) are used to validate the normalizations of the backgrounds. VR1 is defined in the $170 <\KLMbbmm<300$~\GeV\ range, while VR2 is obtained by inverting the \LogL\ selection criterion of SRincl to $-11< \LogL <-8$. All the analysis regions are summarized in Table~\ref{selections} and illustrated in Figure~\ref{regionsPic}.
 
The shapes of the $\ttbar$ kinematic variable  distributions are obtained from simulation, while the overall normalization is extracted from the fits described in Section~\ref{sec:result}.
The distributions for the DY background are taken from data templates because the limited sizes of the simulated event samples do not allow a reliable estimate. The template regions are defined in the same way as the analysis regions in Table~\ref{selections}, except that the two-$b$-tag requirement is replaced by a zero-$b$-tag requirement. The template regions are  ${>}95\%$ pure in DY events. Contributions from other processes, namely \ttbar, $W$+jets, diboson and single-top, are  subtracted using simulation.
Following the subtraction, the DY templates are corrected to account for kinematic differences between event samples dominated by jets originating from light quarks or gluons (template regions) and event samples dominated by $b$-jets (analysis regions). The correction is applied as a per-event weight, where the reweighting is derived from a comparison between two-$b$-tag and zero-$b$-tag kinematic distributions in simulated DY events.
Two sets of event weights are derived and applied sequentially.
First, the jet multiplicity of the zero-$b$-tag MC sample is reweighted to the one in the two-$b$-tag sample. It is the distribution with the largest difference between the zero- and two-$b$-tag samples and was hence corrected first.
Secondly, a BDT-based reweighting is employed to further correct the zero-$b$-tag template kinematics. A BDT is trained  on the zero-$b$-tag versus the two-$b$-tag simulated DY samples. The BDT input consists of kinematic properties and angular distributions of the $b$-jets, muons and the two corresponding $a$-boson candidates, as well as  $\met$ and  $\KLMbbmm$. The ratio of the BDT score distributions obtained for the two-b-tag and zero-b-tag simulated events is then applied as a weight to every event from the zero-b-tag DY template, as a function of its BDT score.
Following the BDT-based reweighting, the \KLMbbmm and $\met$ distributions are corrected by up to 20\%. The DY templates are normalized to data in the fits described in Section~\ref{sec:result}.
 
Minor backgrounds include diboson and single-top-quark production, production of a \ttbar\ pair in association with a vector boson, and $W$ boson production in association with $b$-jets.
The estimation of these minor backgrounds relies purely on simulation normalized to the best available theoretical prediction. The events where a jet is misidentified as a muon are taken into account as follows: non-prompt/misidentified muons in $W$+jets and \ttbar events are included in the analysis on the basis of simulation, any contribution of non-prompt/misidentified muons in the DY+jets component is accounted for by the data template, and the  potential contribution from multijet events is found to be negligible.

\begin{figure}
\centering
\includegraphics[width=0.5\hsize]{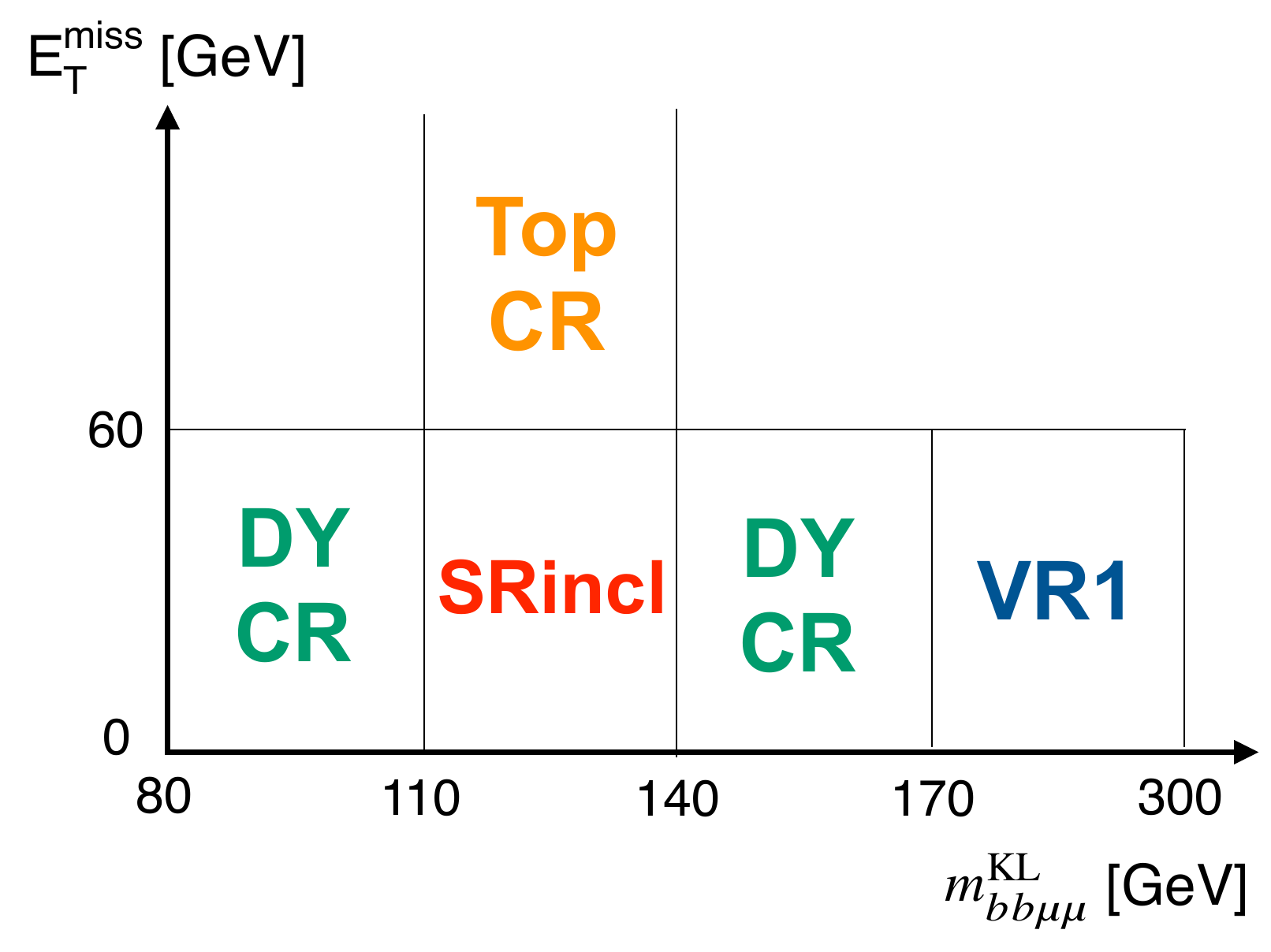}
\caption{Illustration of the signal, control, and validation regions used in the analysis. VR2 (not shown) is defined by the same selection as SRincl, except that the requirement on \LogL\ is inverted to $-11 < \LogL < -8$.}
\label{regionsPic}
\end{figure}
\section{Systematic uncertainties}
Systematic uncertainties in the analysis are divided into three
categories: experimental uncertainties affecting the simulated background and signal processes, uncertainties in the modeling of the DY template, and theoretical uncertainties of the simulated background and signal samples.
Table~\ref{tab:syst} shows a summary of the dominant systematic uncertainties in the total background and signal yields in the signal region bins, as resulting from the fits described in Section~\ref{sec:result} and hereafter denoted by "post-fit".

\begin{table}[!htbp]
\caption{Summary of the dominant post-fit systematic uncertainties in the background and signal yields. The uncertainties are expressed as a percentage of the total background and signal yields per $m_{\mu\mu}$ bin of the signal region.
Only uncertainties exceeding 2\% in at least one SR bin are shown.
\label{tab:syst}}
\begin{center}
\begin{tabular}{l|l|c|c}
\toprule
Category & Source &Total background [\%] & Signal [\%]  \\ \hline
 
\midrule
DY & BDT$m_a$ selection & ~~~~~7--14 &   -- \\
& normalization & ~~~~~5--10 &   -- \\
& \Mmm shape & ~~~1--8 &   -- \\
& kinematics & 0.3--6 &   --  \\
& background subtraction & 0.6--3  &   --  \\
\midrule
$\ttbar$ & hadronization/PS &   0.3--4 & --  \\
& hard-scatter generation  & 0.2--3 & --  \\
& normalization & 0.2--3 & --  \\
\midrule
Overall MC & Sample statistics &  ~~~~~8--40 &  1--2 \\
\midrule
Jets & $b$-tagging &  ~0.03--0.7 &  ~~9--10  \\
& Jet-energy resolution&  ~~~1--3 &  6--7 \\
& Jet-energy scale&  ~~~1--3 &  4--5  \\
\midrule
Signal & FSR & -- &  5 \\
& PS & -- & 4  \\
& VH contribution& -- & ~~~3.5  \\
& MPI & -- & 3  \\
& QCD scale & -- & 3  \\
& ISR & -- &  3 \\
& ggF cross-section & &   \\
& \hspace{1.2cm}- missing higher-order QCD  & -- & 5  \\
& \hspace{1.2cm}- PDF \&  $\alphas$  & -- & 3  \\
\bottomrule
\end{tabular}
\end{center}
\end{table}

Among the experimental uncertainties, the leading effects come from those associated with the
calibration and resolution of jet energies~\cite{PERF-2016-04}, and with the measurement of the $b$-tagging efficiency~\cite{FTAG-2018-01}. The impact of these uncertainties on the total background (signal) yields in the SR bins is as large as 3\% (10\%).
The uncertainty in the combined 2015--2018 integrated luminosity is 1.7\% \cite{ATLAS-CONF-2019-021}, obtained using the LUCID-2 detector \cite{LUCID2} for the primary luminosity measurement.
Other uncertainties, such as those arising from the muon identification efficiency, momentum scale and resolution~\cite{PERF-2015-10,MUON-2018-0333}, and pileup are found to have a negligible impact on the final yields.
 
The uncertainty arising from limited MC sample sizes ranges from 8\% to as large as 40\% in the low $m_a$ mass bins due to there being few \ttbar events in this region.

Five sources of uncertainty in the data-driven DY template are considered. The uncertainty in subtracting non-DY events from the non-reweighted template in the analysis regions is assessed by comparing the nominal template, for which the simulated non-DY backgrounds had been subtracted before reweighting, with an alternative template for which no subtraction had been performed.
The uncertainties in the template kinematics modeling
are derived by comparing the DY template with simulation in two key variables: \met and $\KLMbbmm$.
The ratios of the template to the simulated DY events are fit with linear functions and used in assigning uncertainties to the shapes of the \met and $\KLMbbmm$ distributions.
Similarly, the uncertainty in the  \Mmm template shape is assessed by comparing the template with the smoothed simulated sample and applying the observed difference  as a systematic uncertainty. 
The uncertainty in the normalization of the DY template is obtained from the fits to data.
Finally, the uncertainty in the efficiency of the BDT$m_a$ selection criteria is evaluated by taking the difference in the BDT$m_a$ cut efficiency, $N_{\text{DY events SR}}^{\text{BDTm}\mathrm{_a}\text{>0.2}}/N_{\text{DY events SR}}^{\text{no BDTm}\mathrm{_a} \text{ cut}}$, between the template and the simulation.  All one-sided DY template uncertainties are symmetrized around the nominal value.

To assess the uncertainties in the generation of the hard-scatter \ttbar process, the \textsc{Powheg} sample is compared with a sample generated using \textsc{MadGraph5}\_aMC@NLO~2.3.3. The hadronization and fragmentation uncertainties in the PS are evaluated by comparing the nominal sample showered by \textsc{Pythia~8.230} with an alternative sample generated by \textsc{Powheg} using the same PDF in matrix element as for the nominal sample, but showered with \textsc{Herwig}~7.0.4 \cite{Bahr:2008pv, Bellm:2015jjp}. The initial- and final-state radiation (ISR and FSR) uncertainties of the \ttbar sample are assessed by varying the internal \textsc{Pythia~8.230} showering parameters.
Finally, the uncertainties due to the PDF choice are evaluated using the internal variations of the nominal \textsc{PDF4LHC15\_nlo\_30} set \cite{Butterworth:2015oua}.

Uncertainties in the calculation of the ggF and VBF Higgs boson production cross-sections are assessed by following the recommendations of the LHC Higgs Working Group given in Refs.~\cite{deFlorian:2016spz, Heinemeyer:2013tqa}.
As no VH signal sample was generated, a conservative  100\%  uncertainty is assigned to the estimated VH yield.
To evaluate the  uncertainties due to the PDF choice, the yields obtained with the baseline \textsc{NNPDF30\_nlo\_as\_0118} set are compared with the yields obtained using the internal variations of \textsc{NNPDF30\_nlo\_as\_0118} and with the yields obtained with the nominal \textsc{MMHT2014nlo68clas118} \cite{Harland-Lang:2014zoa} and \textsc{CT14nlo}~\cite{Dulat:2015mca} sets. The largest difference is taken as the overall PDF uncertainty for all signal mass points. Furthermore, the effects of uncertainties in the ISR, FSR, multiparton interactions (MPI) in \textsc{Pythia}, parton showering, and renormalization and factorization scales are also  assessed. Uncertainties from these sources have an impact of 1--6\% on the signal yields, with the largest contributions arising from the ggF production cross-section and FSR uncertainties.

\section{Analysis and results}
\label{sec:result}
The final background and signal estimates are obtained in a set of binned likelihood fits \cite{Cowan:2010js}  using the HistFitter~\cite{Baak:2014wma} package.
The likelihood is a product of Poisson probability functions, describing the observed and predicted numbers of events in each region, and Gaussian distributions that constrain the nuisance parameters associated with the systematic uncertainties.
In the background validation fit, the data in TCR and DYCR are used to extract the normalization of the \ttbar and DY backgrounds, respectively.
As the \ttbar sample in TCR is modeled very well, it is implemented as only one bin in the fit, whereas DYCR is divided into five equal-width bins in \Mmm to provide greater sensitivity to the DY template shape.
The purpose of this fit is to validate the modeling of the background in the control and validation regions and in SRincl. The fitted \ttbar normalization factor is $\mu_{\ttbar}=1.07^{+0.06}_{-0.07}$, while the value of $\mu_\text{DY}$ has no physical meaning because it is scaled from a template region and is thus not quoted. Figures~\ref{SRvalidation1} and \ref{SRvalidation2} show post-fit distributions of \KLMbbmm, \met, \LogL, and \Mmm spanning various analysis regions, while Figure~\ref{SRvalidation3} shows BDT20 and BDT50 in SRincl. Good agreement between the estimated backgrounds and the data is observed in the kinematic distributions.
In SRincl, 1185 events are observed, which is compatible with the total estimated background of $1155.3 \pm 13.6$.
The yields in several representative SR bins, i.e.\ \Mmm windows after applying the BDT selection, as obtained from the background validation fit above, are shown in Table~\ref{SRyields}. When comparing the systematic uncertainty with the statistical uncertainty, it can be seen that the analysis is clearly statistically limited.
Figure~\ref{SRpulls} shows the data and the estimated backgounds in all final SR bins. Due to the limited statistics of the background samples, the estimates are not perfectly smooth; however, the bin-to-bin fluctuations are much smaller than the statistical uncertainty of the data. Larger jumps, which occur at $m_a = 23$, 28, 33, 38~\GeV\ etc., appear when the BDT discriminant used for the selection changes from the one trained in the lower mass range to the one trained in the higher mass range.

\begin{table}[htb]
\caption{Total and individual background yields in six
representative $m_{\mu\mu}$ bins of the signal region after the BDT selection is applied. The yields
are the post-fit values as determined by the background validation fit.
The uncertainties shown include all systematic and statistical uncertainties. As the diboson, single top quark, $\ttbar V$, and $W$+jets contributions are very small, they are summed in the table under "Other".}
\begin{center}
\setlength{\tabcolsep}{0.5pc}{\small
\begin{tabular*}{\textwidth}{@{\extracolsep{\fill}}l|c|c|c|c|c|c}
\toprule
$m_{\mu\mu}$ bin [\GeV] &  [15--17]   & [24--26]   & [34--36]   & [44--46] &  [50.5--53.5]    & [60.5--63.5]               \\[-0.05cm]
\midrule
\midrule
Observed events          & $6$              & $9$              & $19$              & $17$              & $39$              & $8$                    \\
\noalign{\smallskip}\hline\noalign{\smallskip}
Total background         & $4.8 \pm 2.2$          & $9.0 \pm 1.8$          & $11.9 \pm 1.6$~~          & $15.5 \pm 2.0$~~          & $19.3 \pm 2.7$~~          & $9.3 \pm 1.7$              \\
\noalign{\smallskip}\hline\noalign{\smallskip}
DY         & $4.6 \pm 2.1$          & $6.4 \pm 1.5$          & $5.7 \pm 1.1$          & $6.4 \pm 1.5$          & $8.3 \pm 2.1$          & $5.3 \pm 1.4$              \\
\ttbar         & $0.2 \pm 0.1$          & $2.6 \pm 0.8$          & $6.0 \pm 1.1$          & $8.5 \pm 1.4$          & $10.4 \pm 2.4$~~          & $3.5 \pm 0.9$              \\
Other         & $0.03 \pm 0.01$          &   $0.03 \pm 0.00$          &    $0.24 \pm 0.12$          & $0.50 \pm 0.40$           & $0.50 \pm 0.12$          & $0.45 \pm 0.19$             \\
\bottomrule
\end{tabular*}
}
\end{center}
\label{SRyields}
\end{table}

\begin{figure}
\centering
\includegraphics[width=0.48\hsize]{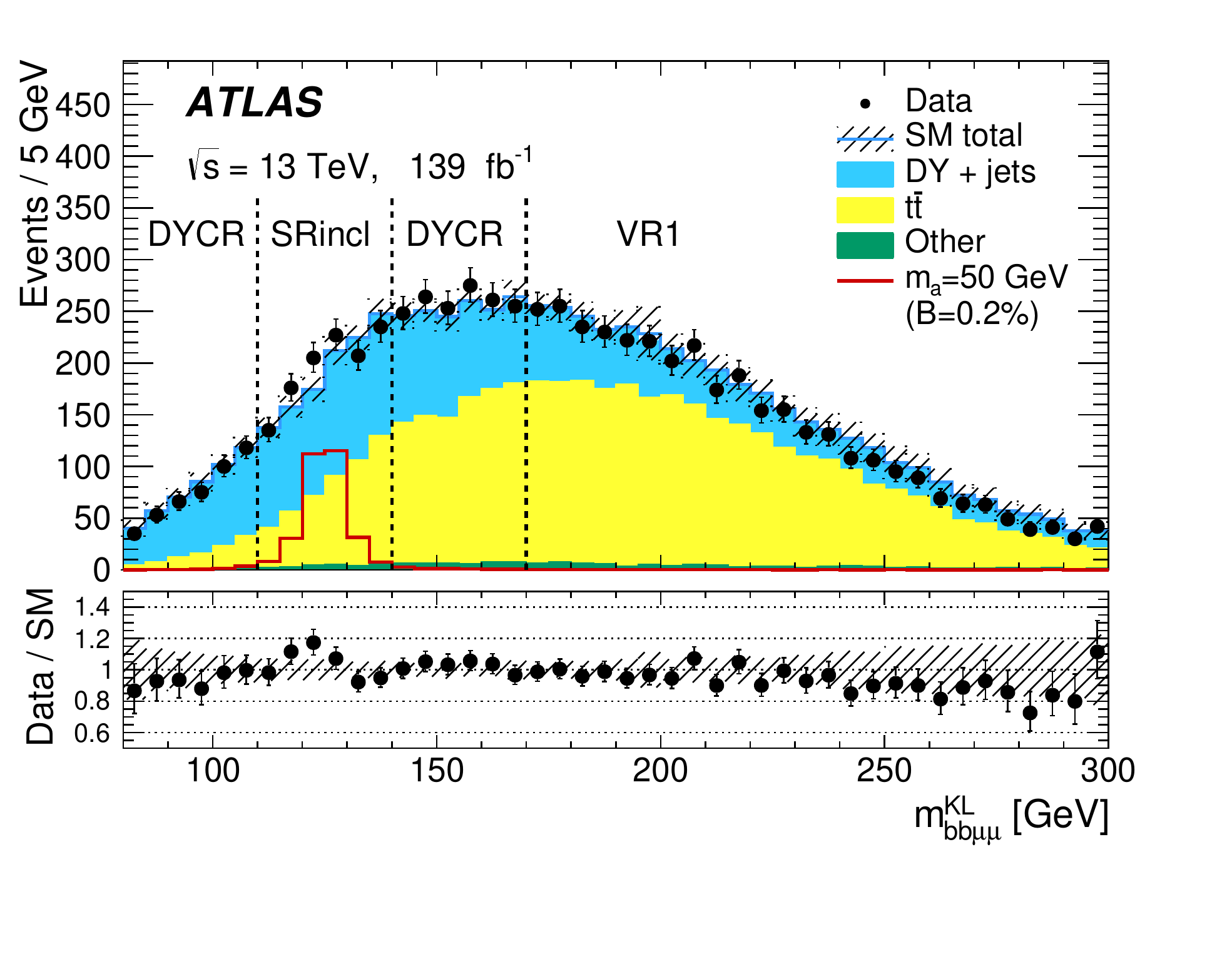}
\includegraphics[width=0.48\hsize]{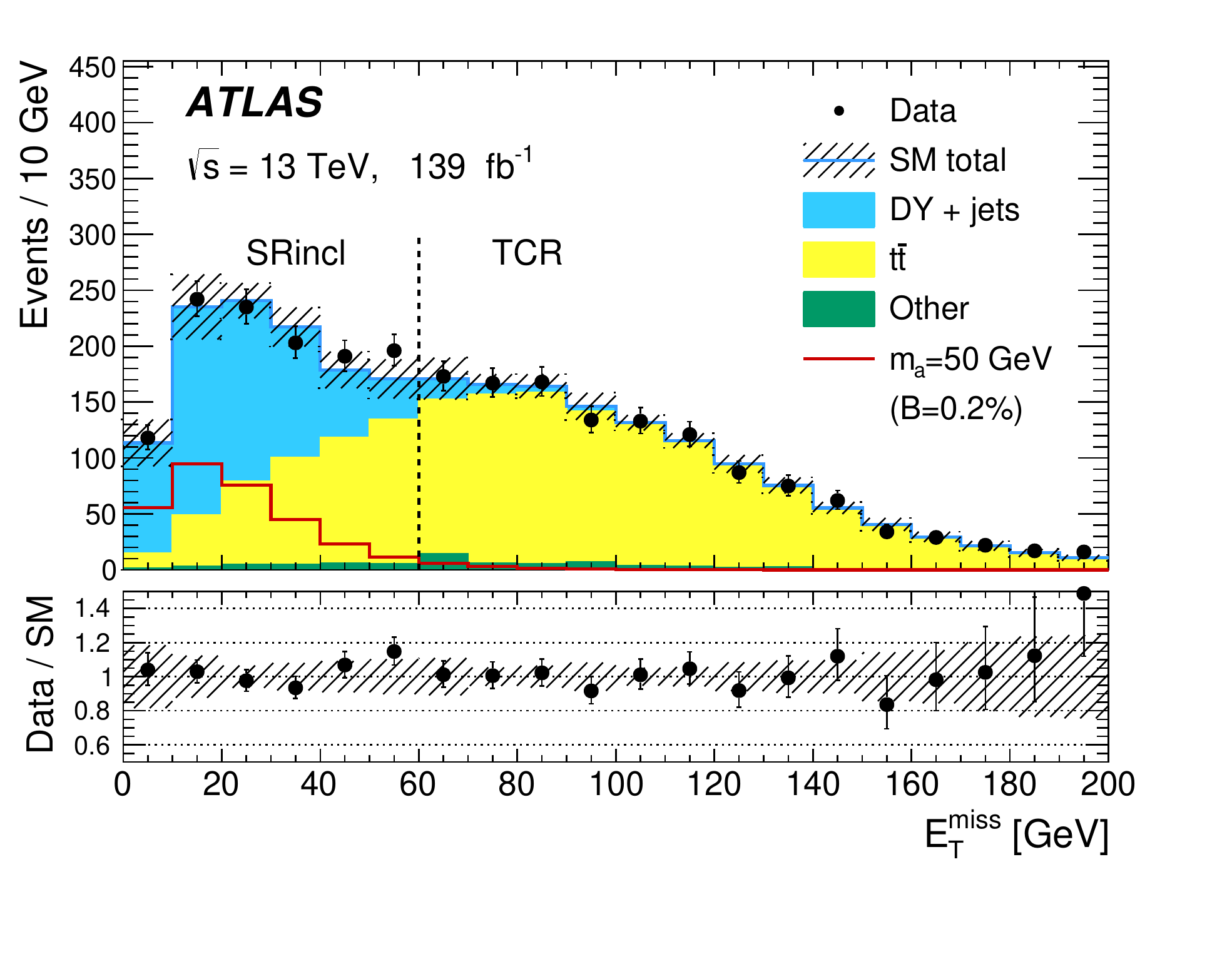}
\caption{Post-fit \KLMbbmm in DYCR, SRincl and VR1 (left); \met in SRincl and TCR (right). No selection based on the BDT discriminants is applied in the analysis regions shown in the figures. The signal distributions are normalized to the SM Higgs boson cross-section (including ggF, VBF, and VH production) and assume
$\mathcal{B}(H \rightarrow aa \rightarrow bb\mu\mu)$ as indicated in the legends of the figures (chosen to ensure good visibility in the plot). The hatched bands show the total post-fit statistical and systematic uncertainties of the backgrounds. The histogram labeled as ``Other'' in the legend includes the contributions from the diboson, single-top-quark, $\ttbar{+}V$ and $W$+jets backgrounds.}
\label{SRvalidation1}
\end{figure}
 
\begin{figure}
\centering
\includegraphics[width=0.48\hsize]{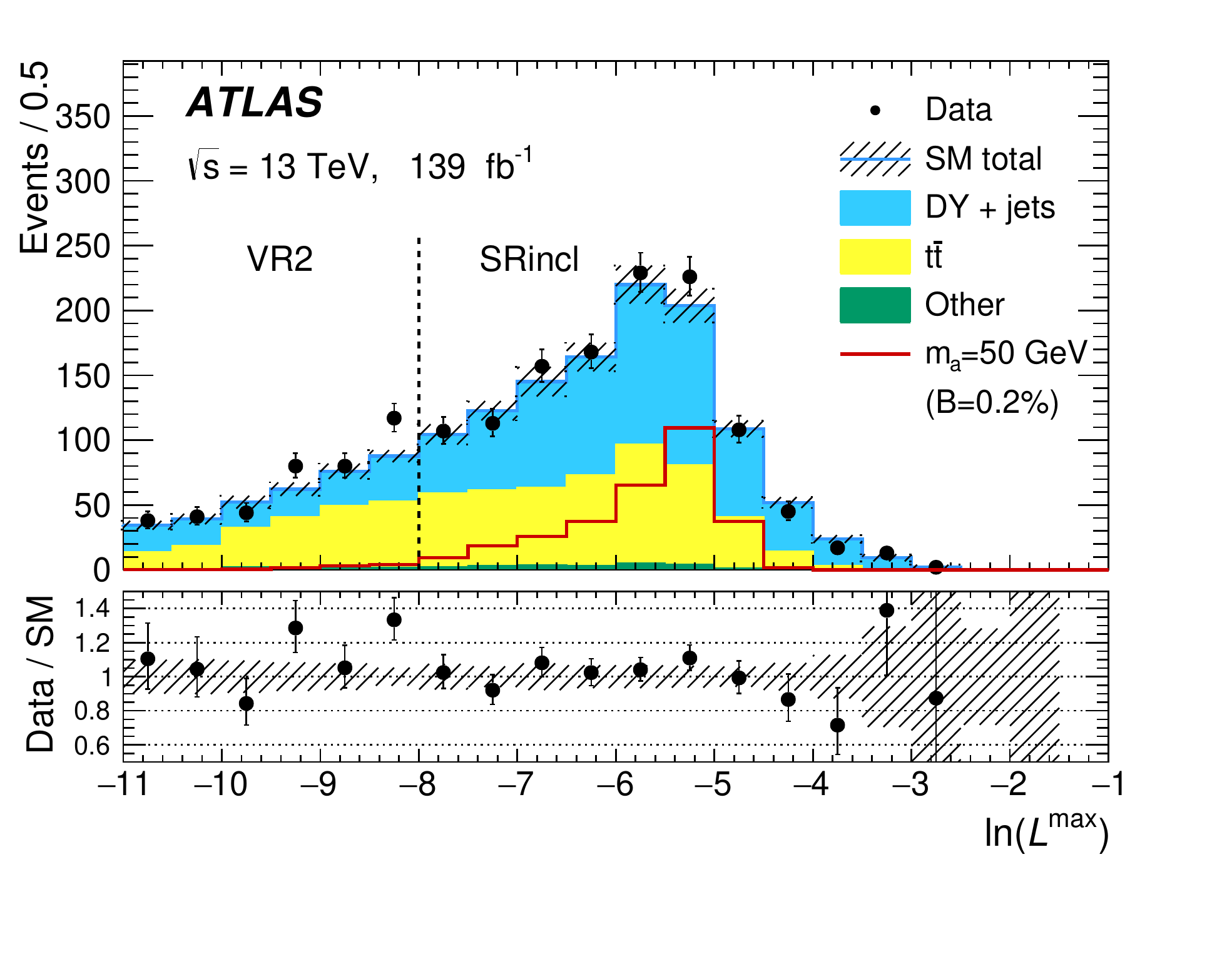}
\includegraphics[width=0.48\hsize]{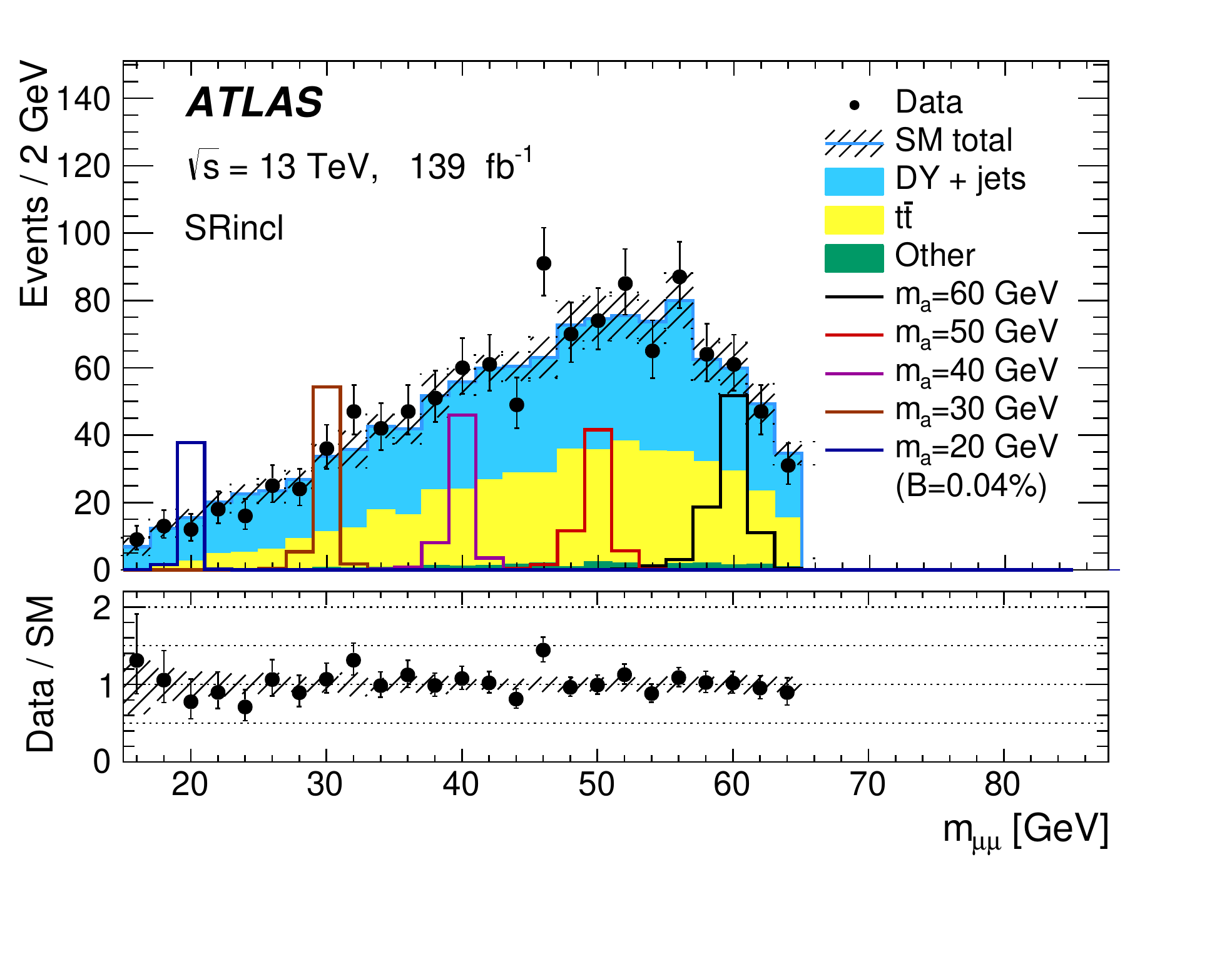}
\caption{Post-fit \LogL~in VR2 and SRincl (left); \Mmm in SRincl (right). No selection based on the BDT discriminants is applied in the analysis regions shown in the figures. The signal distributions are normalized to the SM Higgs boson cross-section (including ggF, VBF, and VH production) and assume
$\mathcal{B}(H \rightarrow aa \rightarrow bb\mu\mu)$ as indicated in the legends of the figures (chosen to ensure good visibility in the plot). The hatched bands show the total post-fit statistical and systematic uncertainties of the backgrounds. The histogram labeled as ``Other'' in the legend includes the contributions from the diboson, single-top-quark, $\ttbar{+}V$ and $W$+jets backgrounds. }
\label{SRvalidation2}
\end{figure}

\begin{figure}
\centering
\includegraphics[width=0.48\hsize]{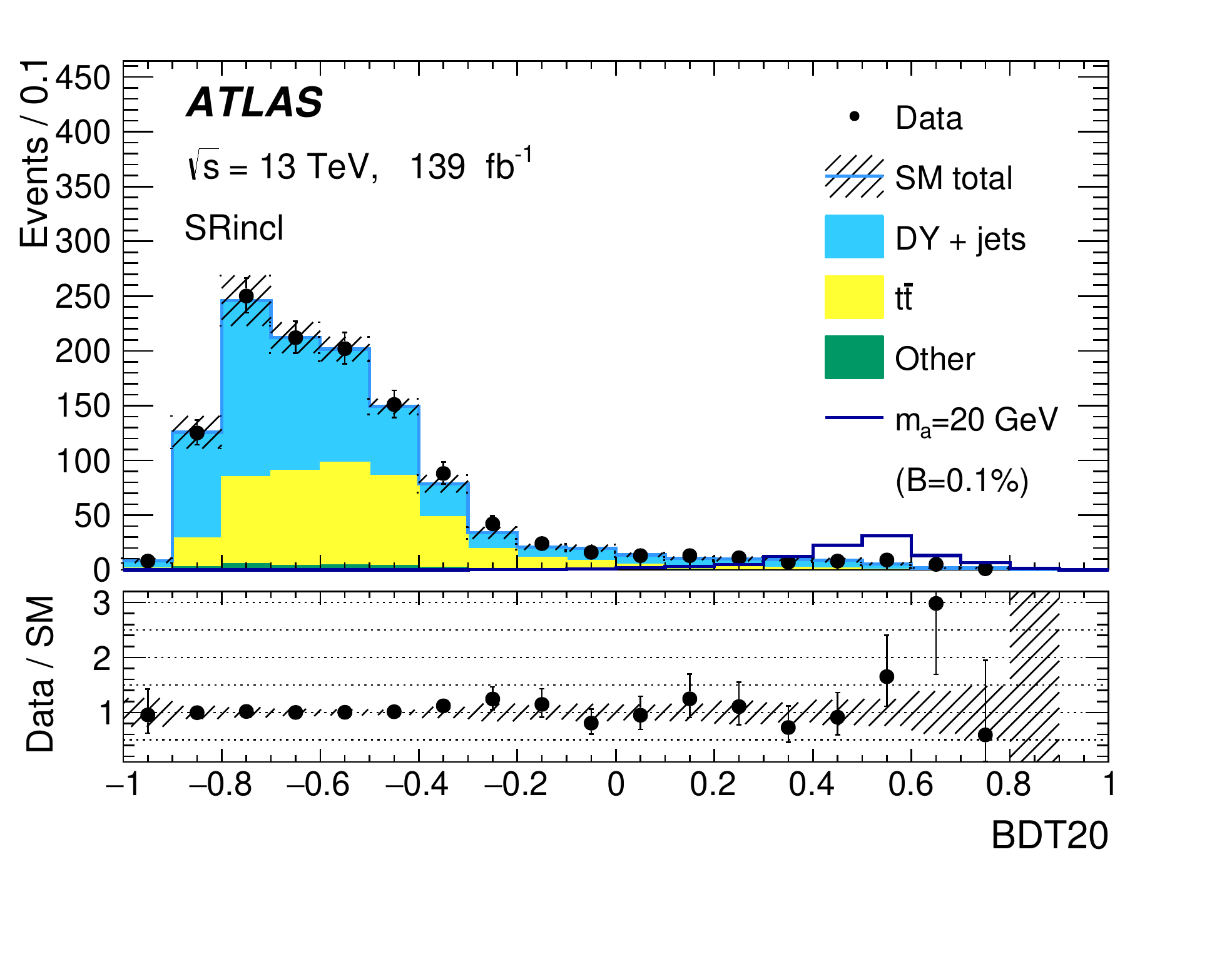}
\includegraphics[width=0.48\hsize]{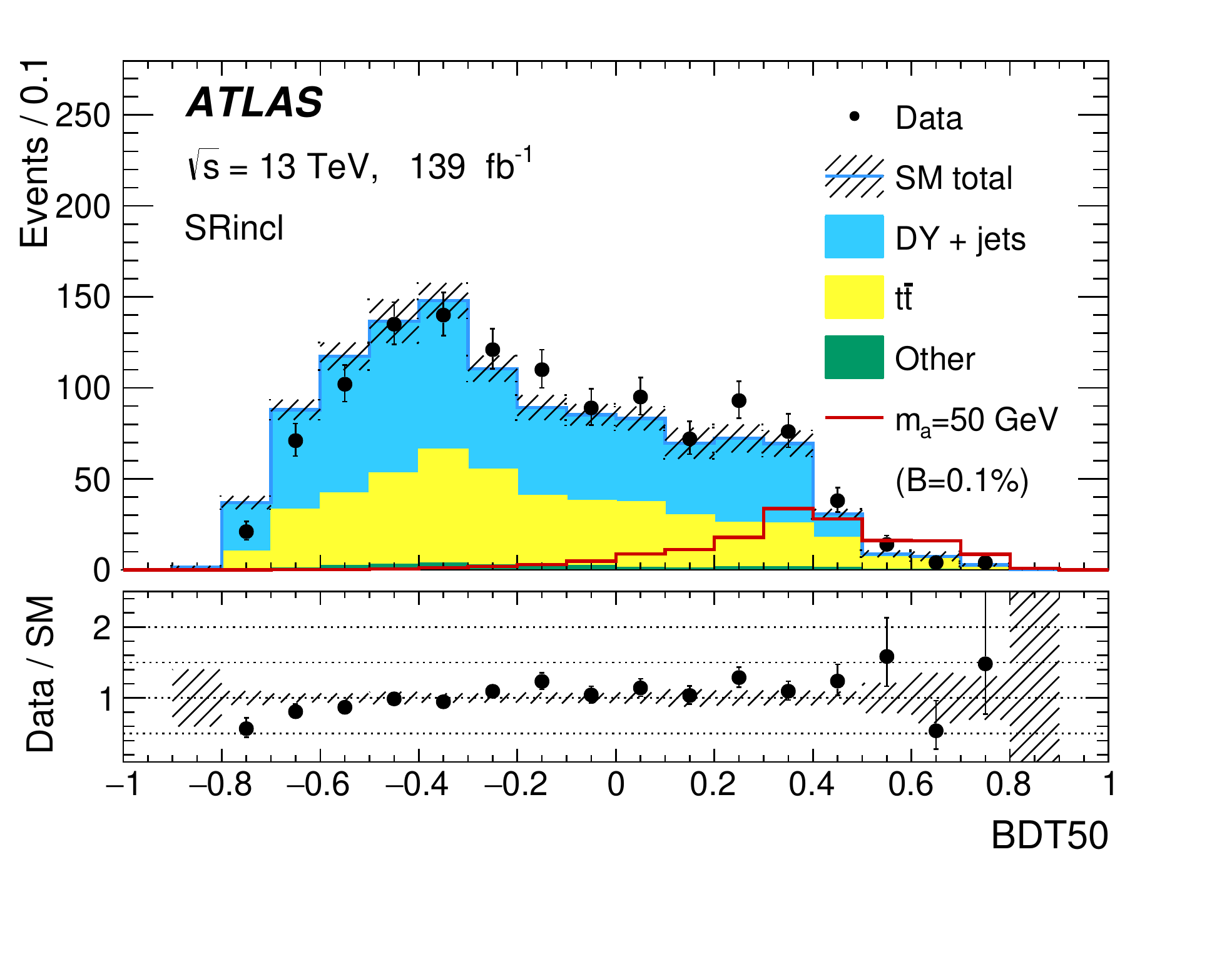}
\caption{Post-fit BDT20 (left) and BDT50 (right) distributions in SRincl. The signal distributions are normalized to the SM Higgs boson cross-section (including ggF, VBF, and VH production) and assume
$\mathcal{B}(H \rightarrow aa \rightarrow bb\mu\mu)$ as indicated in the legends of the figures (chosen to ensure good visibility in the plot). The hatched bands show the total post-fit statistical and systematic uncertainties of the backgrounds. The histogram labeled as ``Other'' in the legend includes the contributions from the diboson, single-top-quark, $\ttbar{+}V$ and $W$+jets backgrounds. }
\label{SRvalidation3}
\end{figure}

\begin{figure}
\centering
\includegraphics[width=0.99\hsize]{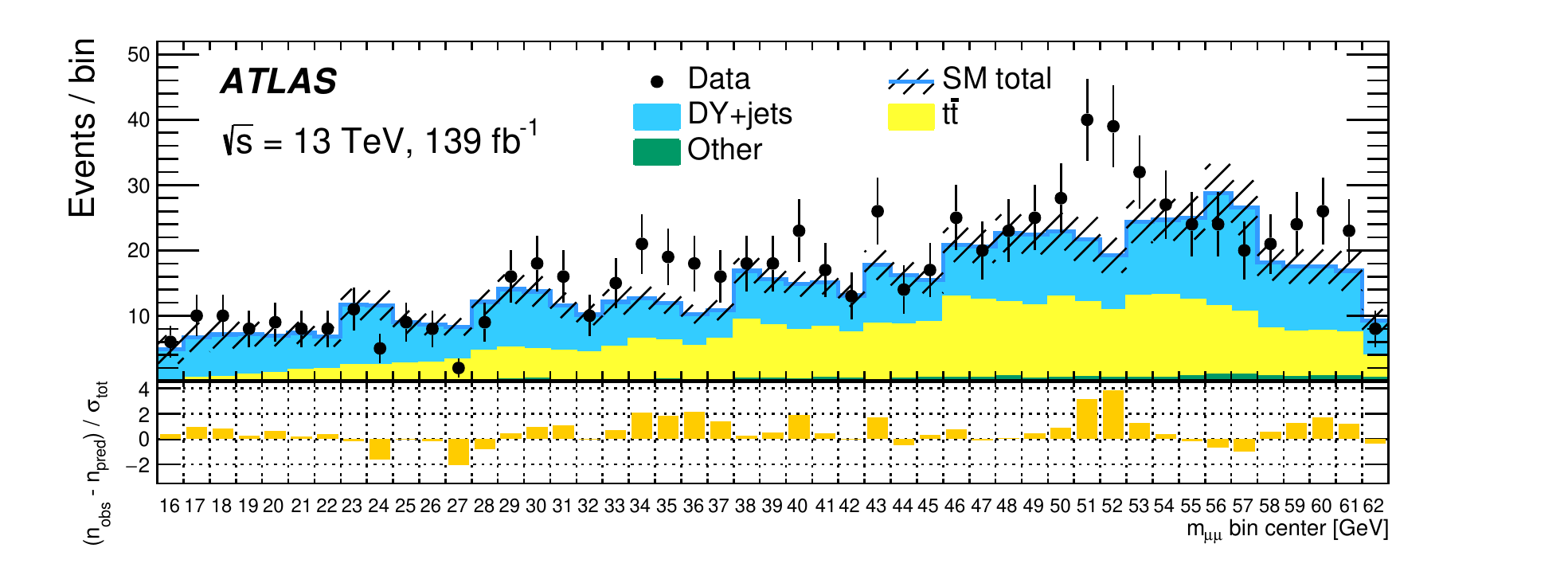}
\caption{Post-background-validation-fit number of  events in all SR bins (after applying the BDT selection) that are tested for the presence of signal. The bin widths are 2~\GeV\ (3~\GeV) in \Mmm for $m_a\leq45$~\GeV\ ($m_a>45$~\GeV). Neighboring bins partially overlap, hence they are not statistically independent.
The bottom panel shows the pull in each bin, defined as $(n_\mathrm{obs}-n_\mathrm{pred})/\sigma_\mathrm{tot}$, where $n_\mathrm{obs}$ is the number of events in the data, $n_\mathrm{pred}$ is the number of fitted background events and $\sigma_\mathrm{tot}$ is the total (systematic and statistical, added in quadrature) uncertainty in the fitted background yield.
Discontinuities in the background predictions appear when the BDT discriminant used for the selection changes from the one trained in the lower mass range to the one trained in the higher mass range. The histogram labeled as ``Other'' in the legend includes the contributions from the diboson, single-top-quark, $\ttbar{+}V$ and $W$+jets backgrounds.}
\label{SRpulls}
\end{figure}

To test for the presence of new phenomena, fits are performed for each of the 47 hypothesized signal masses in the range $16\leq\Mmm\leq62$~\GeV\ in 1~\GeV\ steps.
It was verified that the analysis is also sufficiently sensitive to a signal with \Mmm centered in between these 1~\GeV\ steps.
TCR, DYCR, and the respective SR bin are included in each fit in order to constrain  the backgrounds and the signal to the data.

A model-independent fit, i.e.\ not including any signal sample, is performed to test whether the data are compatible with the background-only hypothesis. The result  is a scan of $p_0$-values as shown in Figure~\ref{pvalue}. The largest discrepancy is found at $\Mmm=52$~\GeV, corresponding to a local (global) $p_0$-value of 0.00054 (0.048) and a local (global) significance of $3.3 \sigma$ ($1.7 \sigma$). The global significance was calculated from the asymptotic formulae in Refs.~\cite{Cowan:2010js,Gross_2010}.

Upper limits, derived using  the CL$_\text{s}$ technique \cite{Read:2002hq,Junk:1999kv}, are set on $\mathcal{B}(H \rightarrow aa \rightarrow bb\mu\mu)$ in a series of conditional fits, this time also including the signal samples.
The limits as a function of $m_a$ are shown in Figure~\ref{Limits}.
Uniform sensitivity is achieved for all masses above 18~\GeV, while for lower signal masses, $m_a \leq 18$~\GeV, the sensitivity of the analysis decreases due to $b$-jets falling below the reconstruction threshold or merging into one reconstructed jet.
Figure~\ref{SRwithSign1} shows \Mmm and BDT$m_a$ distributions  after the signal+background fit for two SR bins, $m_a=35$~\GeV\ and $m_a=52$~\GeV, where the two largest deviations from the background-only hypothesis are observed.
The signal in the plots is scaled to the best-fit value, corresponding to $\mathcal{B}(H \rightarrow aa \rightarrow bb\mu\mu)=6.4\times 10^{-5}$ ($1.9\times 10^{-4}$) for $m_a=35$~\GeV\ ($m_a=52$~\GeV).
 
The upper limits at 95\% CL on $\mathcal{B}(H\rightarrow aa\rightarrow bb\mu\mu)$ range between $0.2 \times 10^{-4}$ and $4.0 \times 10^{-4}$, depending on $m_a$.
These limits improve upon the previous ATLAS result based on 36~$\ifb$ of data~\cite{bbmm36ifb} by a factor of 2--5 over the full  \Mmm range. A factor of $\sim$2 improvement in sensitivity comes from the larger dataset, and a further factor of $\sim$2 is achieved thanks to the use of multivariate techniques to discriminate between the signal and the SM backgrounds. Due to small number of background events at lower signal masses $m_a$, the BDT training is less efficient in this region, and the gain from applying the BDT$m_a$ selection criteria is higher at higher $m_a$.
Taking as an example the favorable scenario with $\mathcal{B}(H\rightarrow aa\rightarrow bb\mu\mu)/ \mathcal{B}(H\rightarrow aa )= 0.16\%$, the analysis probes the Higgs boson branching fraction into pseudoscalars down to $\mathcal{B}(H\rightarrow aa) = 1.3\%$, much lower than the limits derived from combinations of the Higgs boson measurements.
 
So as not to restrict the analysis sensitivity solely to models where the $a$-particle is a pseudoscalar, upper limits obtained without employing the BDT discriminants are also derived as shown in Figure~\ref{Limits_noBDT}.
In addition to being less sensitive to the particle's CP properties, the limits in SRincl without the BDT selection also facilitate reinterpretations of the analysis.
These limits are derived in the same way as described above, i.e.\ by scanning the \Mmm windows of SRincl, but omitting the final selection on the BDT discriminants. The expected limits obtained when employing the baseline analysis strategy are also shown in Figure~\ref{Limits_noBDT} for comparison, illustrating the significant improvement in sensitivity to pseudoscalars when using the BDTs. The excess observed at $\Mmm=52$~\GeV\ in the BDT analysis is not supported by the limits derived without the BDTs.
Figure~\ref{SRpullsnoBDT} shows the data and the estimated backgrounds in all final SR bins, without applying the BDT selection.

\begin{figure}
\centering
\includegraphics[width=0.6\hsize]{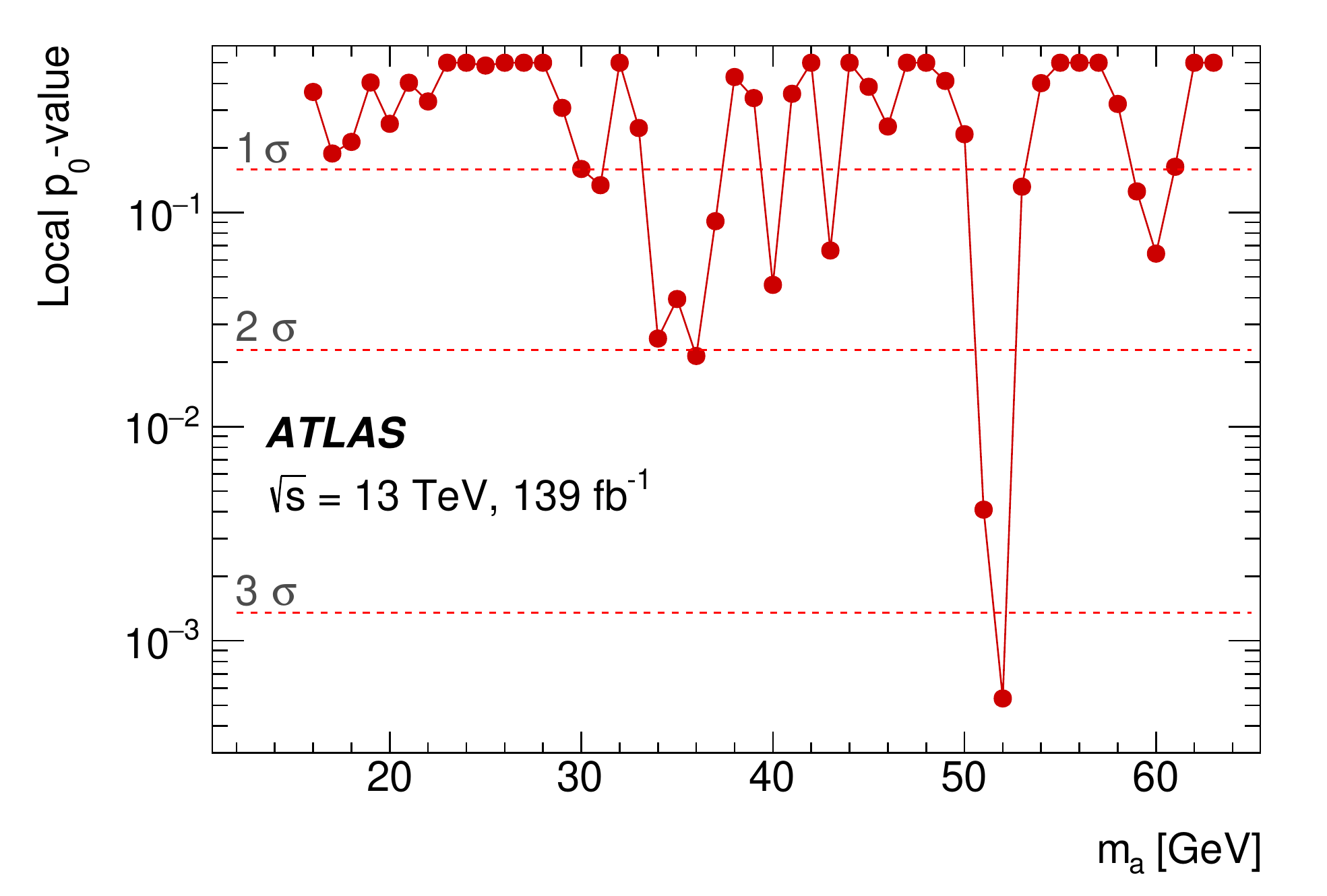}
\caption{The local $p_0$-values are quantified in standard deviations $\sigma$ and plotted as a function of the signal mass hypothesis. Between the points, the $p_0$-values are interpolated and may not be fully representative of the actual sensitivity.}
\label{pvalue}
\end{figure}

\begin{figure}
\centering
\includegraphics[width=0.6\hsize]{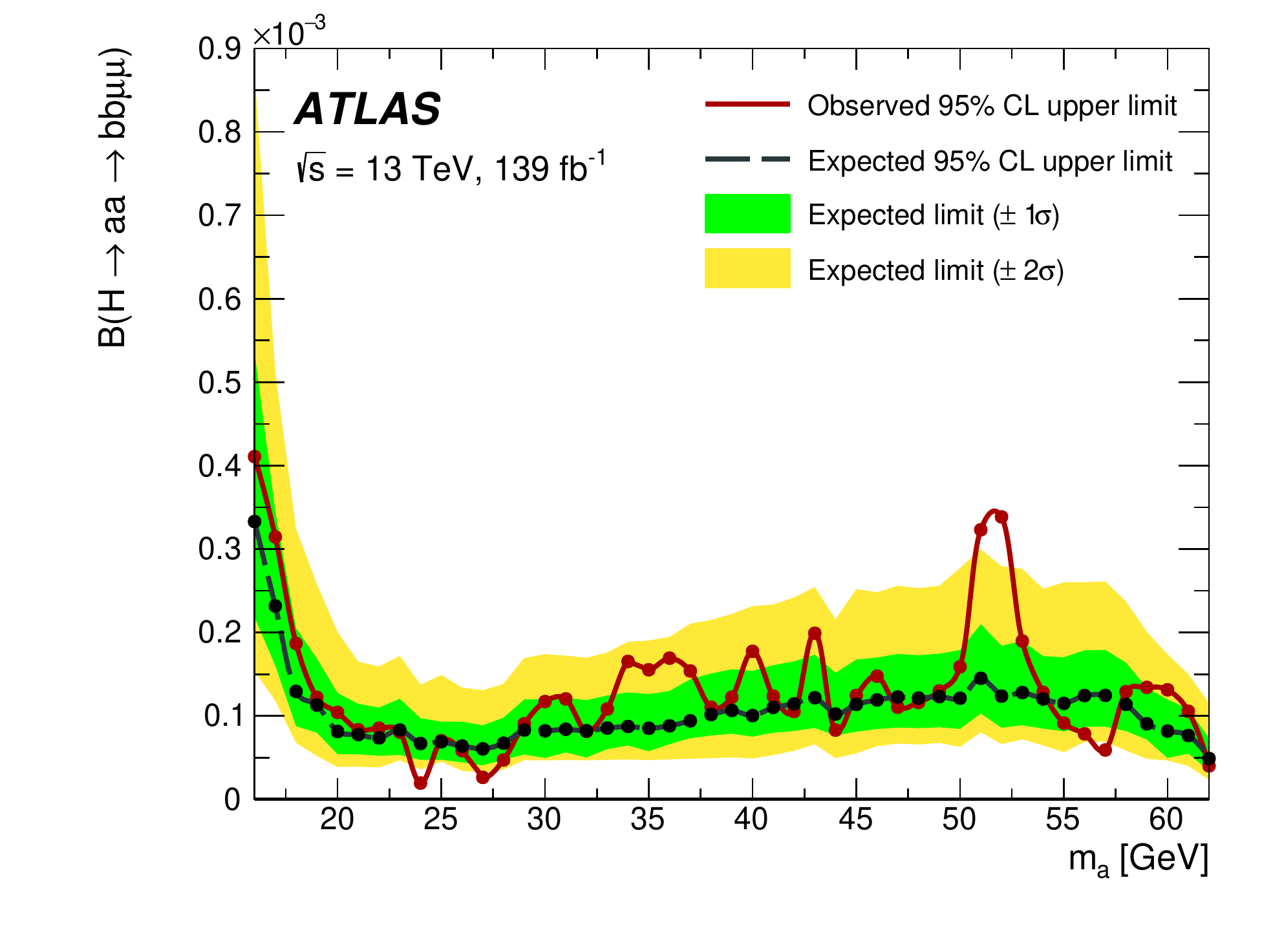}
\caption{Upper limits on $\mathcal{B}(H \rightarrow aa \rightarrow bb\mu\mu)$ at 95\% CL, including the BDT selection, as a function of the signal mass hypothesis. Black and red dots show masses for which the hypothesis testing was done. Between these points, the limits are interpolated and may not be fully representative of the actual sensitivity.}
\label{Limits}
\end{figure}

\begin{figure}
\centering
\includegraphics[width=0.48\hsize]{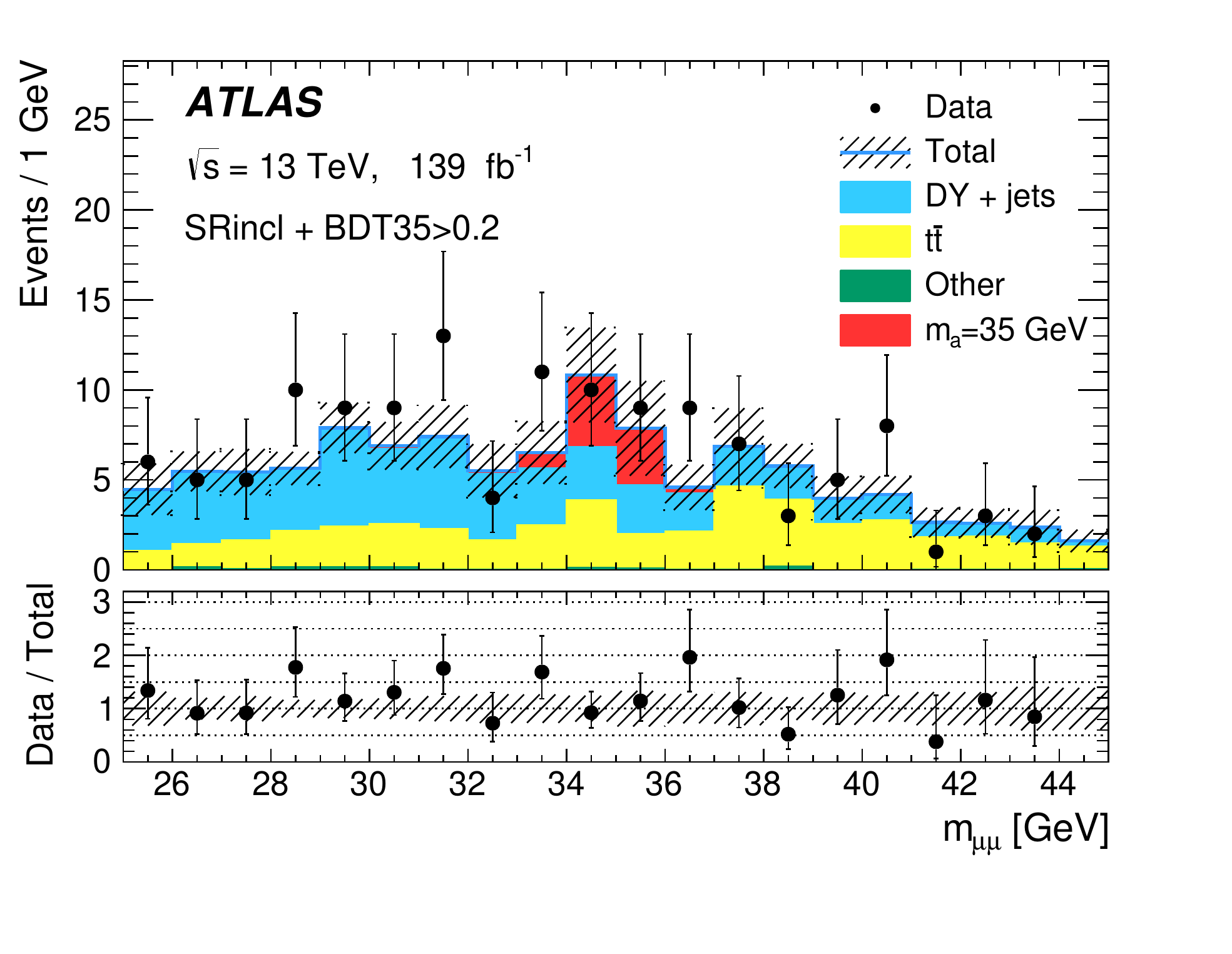}
\includegraphics[width=0.48\hsize]{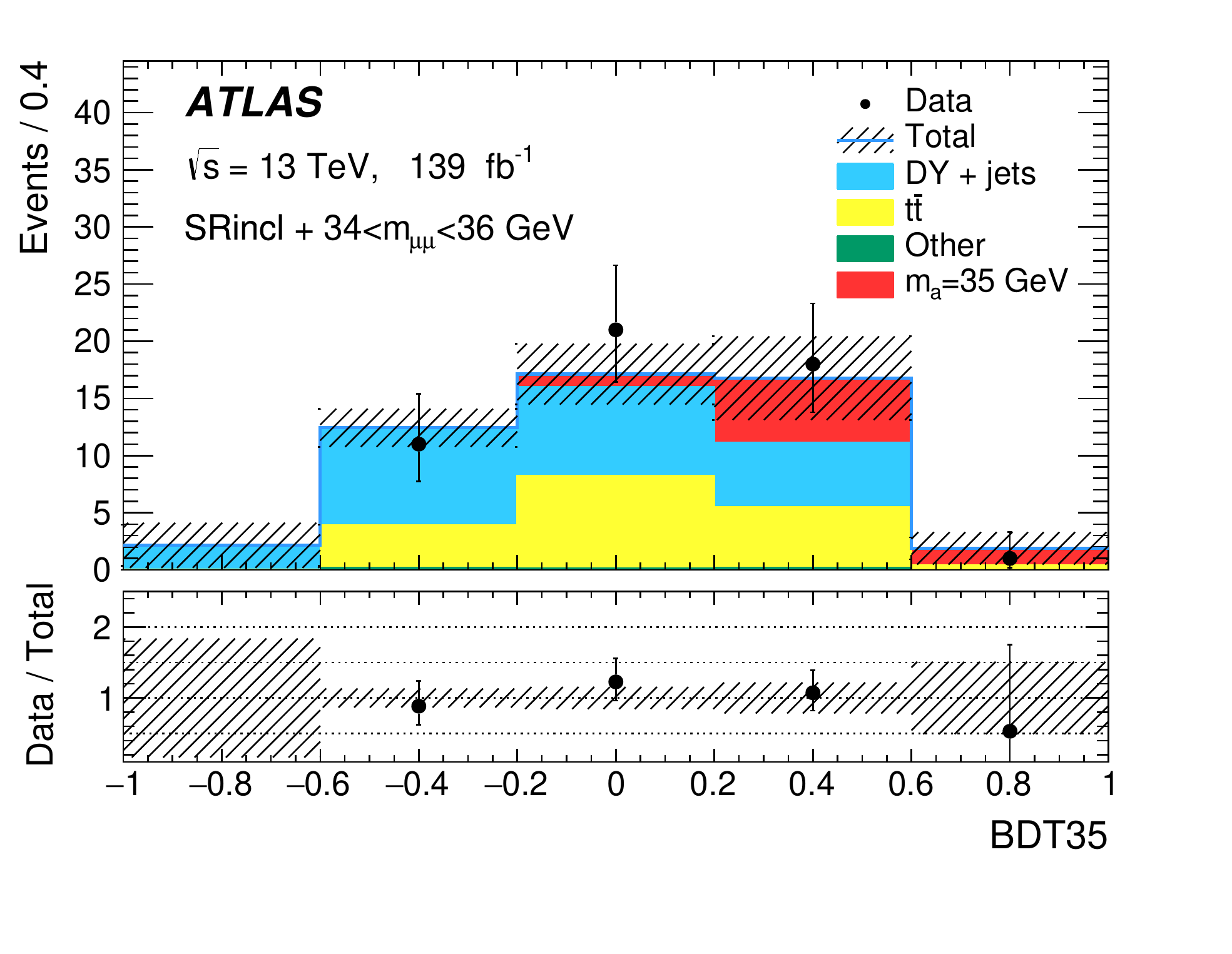}
\includegraphics[width=0.48\hsize]{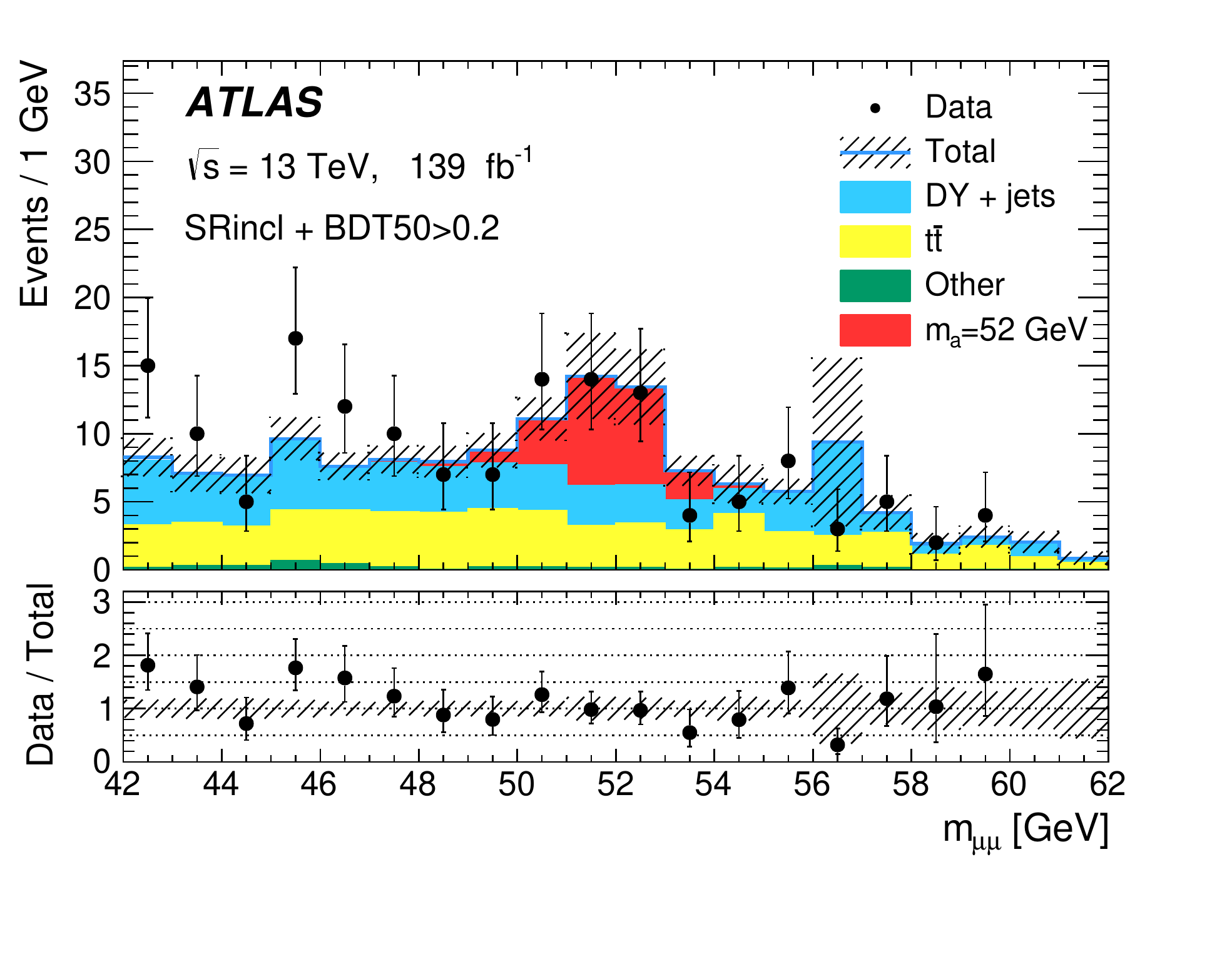}
\includegraphics[width=0.48\hsize]{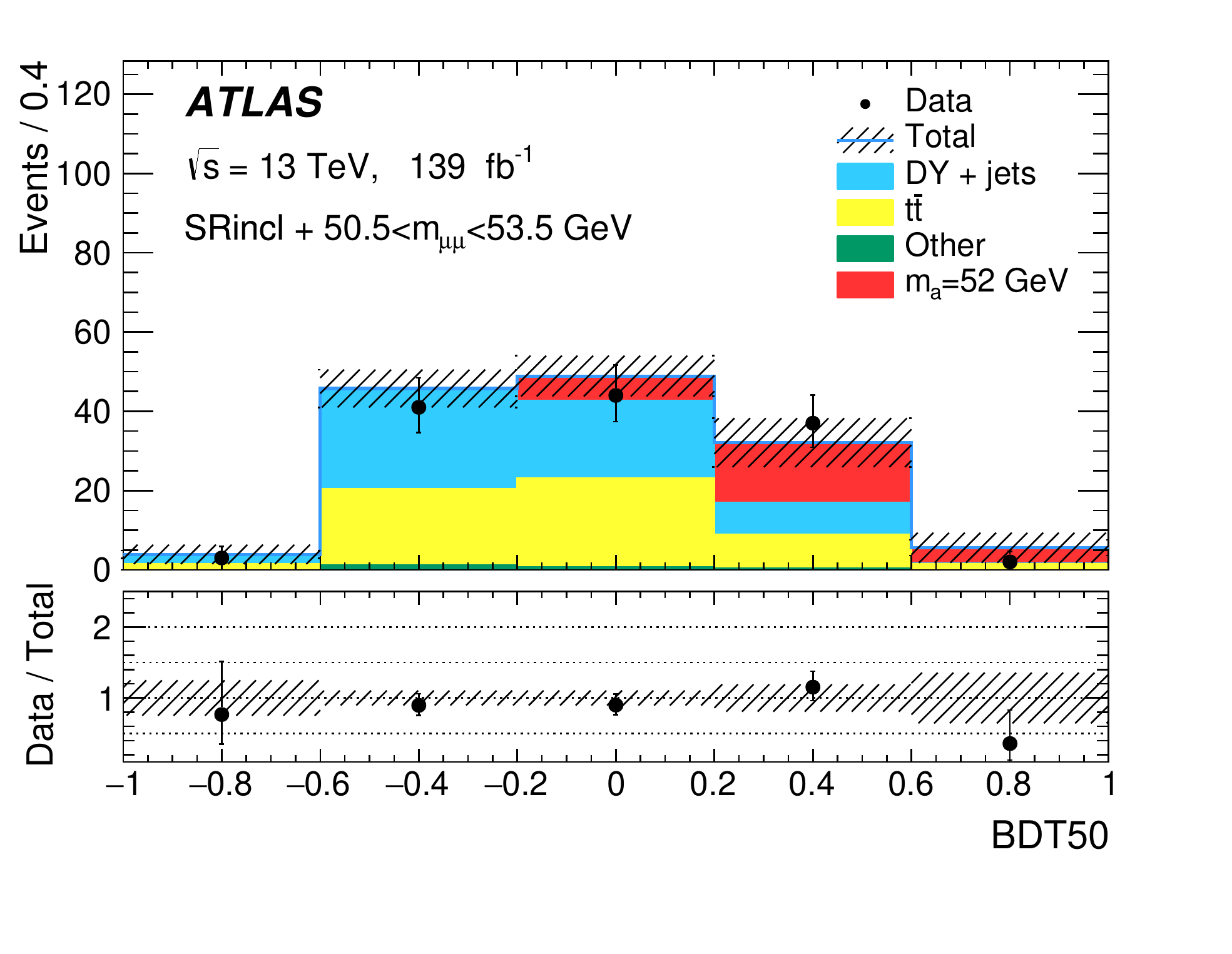}
\caption{\Mmm distributions in the SRincl after the BDT$35>0.2$ selection (top left) and BDT$50>0.2$ selection (bottom left), and BDT35 (top right) and BDT50 (bottom right) distributions in the SRincl in the \Mmm window 34--36~\GeV\ and 50.5--53.5~\GeV, respectively.
The signal is scaled to the best-fit value, $\mathcal{B}(H \rightarrow aa \rightarrow bb\mu\mu)=6.4\times 10^{-5}$ for the top plots, and $1.9\times 10^{-4}$ for the bottom plots, assuming the SM Higgs boson cross-section (including ggF, VBF, and VH production). The hatched bands show the total post-fit statistical and systematic uncertainties of the backgrounds and the signal. The histogram labeled as ``Other'' in the legend includes the contributions from the diboson, single-top-quark, $\ttbar{+}V$ and $W$+jets backgrounds.}
\label{SRwithSign1}
\end{figure}

\begin{figure}
\centering
\includegraphics[width=0.6\hsize]{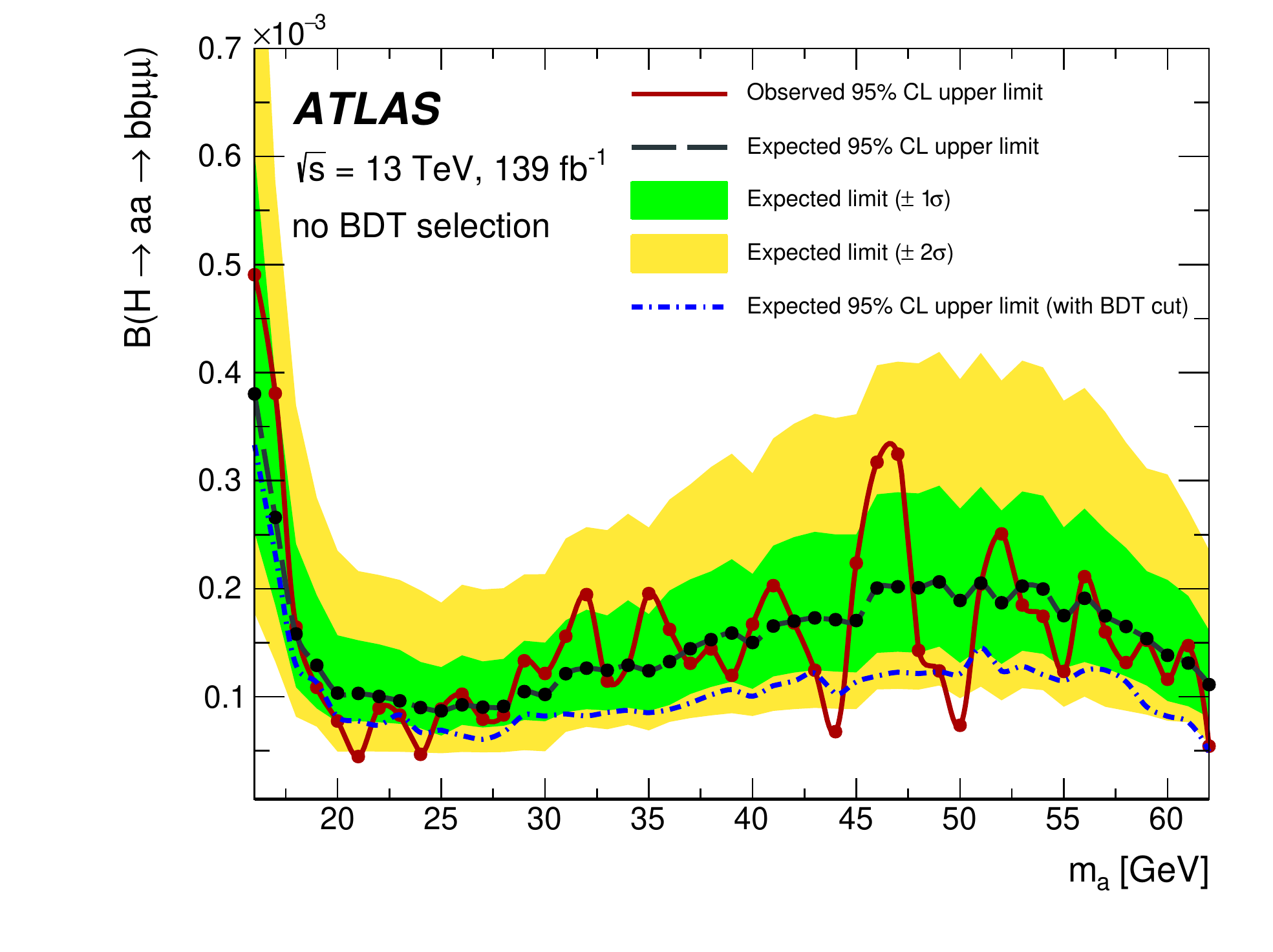}
\caption{Upper limits on $\mathcal{B}(H \rightarrow aa \rightarrow bb\mu\mu)$  at 95\% CL, with no BDT selection applied, as a function of the signal mass hypothesis. The dash-dotted blue line indicates the expected limit set in the analysis with the BDT selection. Black and red dots show masses for which the hypothesis testing was done. Between these points, the limits are interpolated and may not be fully representative of the actual sensitivity. }
\label{Limits_noBDT}
\end{figure}

\begin{figure}
\centering
\includegraphics[width=0.99\hsize]{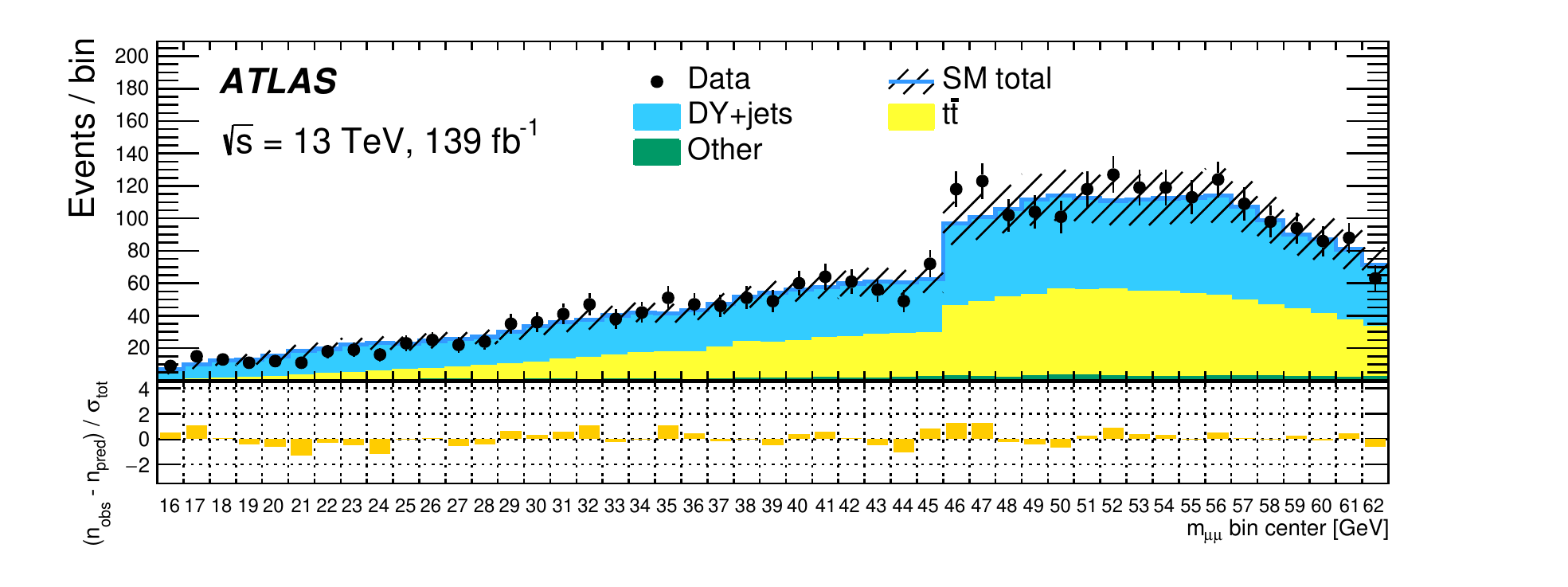}
\caption{Post-background-validation-fit number of  events in all SR bins (without applying the BDT selection) that are tested for the presence of signal. The bin widths are 2~\GeV\ (3~\GeV) in \Mmm for $m_a\leq45$~\GeV\ ($m_a>45$~\GeV). Neighboring bins partially overlap, hence they are not statistically independent.
The bottom panel shows the pull in each bin, defined as $(n_\mathrm{obs}-n_\mathrm{pred})/\sigma_\mathrm{tot}$, where $n_\mathrm{obs}$ is the number of events in the data, $n_\mathrm{pred}$ is the number of fitted background events and $\sigma_\mathrm{tot}$ is the total (systematic and statistical, added in quadrature) uncertainty in the fitted background yield. The discontinuity at $m_a=45$~\GeV\ appears where the \Mmm window size is changed. The histogram labeled as ``Other'' in the legend includes the contributions from the diboson, single-top-quark, $\ttbar{+}V$ and $W$+jets backgrounds.}
\label{SRpullsnoBDT}
\end{figure}

\FloatBarrier
 
 
\section{Conclusion}
 
A search for light pseudoscalar particles (denoted by $a$) in the decays of the 125~\GeV\ Higgs boson in the final state with two muons and two $b$-tagged jets, $H \rightarrow aa \rightarrow bb\mu\mu$, is presented.
The analysis is performed using 139~fb$^{-1}$ of $\sqrt{s}=13$~\TeV\ $pp$ collision data recorded by the ATLAS detector at the LHC between 2015 and 2018. A narrow resonance is searched for in the dimuon invariant mass spectrum in the range $16 \leq \Mmm \leq 62$~\GeV.
BDT classifiers are trained to distinguish the $H \rightarrow aa$ signal, where $a$ is a pseudoscalar, from the SM backgrounds.
Additionally, the result without selection on the BDT discriminants is also provided to ensure sensitivity to models where the $a$-particle is not necessarily a pseudoscalar, as well as  to facilitate reinterpretations of the analysis.
No significant excess of the data above the SM backgrounds is observed.
In the BDT analysis, the lowest local $p_0$-value of 0.00054 is observed at $\Mmm= 52$~\GeV\ and corresponds to a local significance of $3.3 \sigma$. The global significance of that excess is determined to be $1.7 \sigma$.
Upper limits at 95\%~CL including (excluding) the BDT selection criteria are set on $\mathcal{B}(H \rightarrow aa \rightarrow bb\mu\mu)$ and range between $0.2\times10^{-4}$ and $4.0\times10^{-4}$ ($0.5\times10^{-4}$ and $5.0\times10^{-4}$), depending on $m_a$.
The result including the BDT selection criteria improves upon previous ATLAS and CMS limits by about a factor of 2--5 for $m_a>20$~\GeV, while both results (with and without the BDT) extend the search down to $m_a$ values of 16~\GeV.

 
\section*{Acknowledgments}
 

We thank CERN for the very successful operation of the LHC, as well as the
support staff from our institutions without whom ATLAS could not be
operated efficiently.
 
We acknowledge the support of
ANPCyT, Argentina;
YerPhI, Armenia;
ARC, Australia;
BMWFW and FWF, Austria;
ANAS, Azerbaijan;
SSTC, Belarus;
CNPq and FAPESP, Brazil;
NSERC, NRC and CFI, Canada;
CERN;
ANID, Chile;
CAS, MOST and NSFC, China;
Minciencias, Colombia;
MSMT CR, MPO CR and VSC CR, Czech Republic;
DNRF and DNSRC, Denmark;
IN2P3-CNRS and CEA-DRF/IRFU, France;
SRNSFG, Georgia;
BMBF, HGF and MPG, Germany;
GSRI, Greece;
RGC and Hong Kong SAR, China;
ISF and Benoziyo Center, Israel;
INFN, Italy;
MEXT and JSPS, Japan;
CNRST, Morocco;
NWO, Netherlands;
RCN, Norway;
MEiN, Poland;
FCT, Portugal;
MNE/IFA, Romania;
JINR;
MES of Russia and NRC KI, Russian Federation;
MESTD, Serbia;
MSSR, Slovakia;
ARRS and MIZ\v{S}, Slovenia;
DSI/NRF, South Africa;
MICINN, Spain;
SRC and Wallenberg Foundation, Sweden;
SERI, SNSF and Cantons of Bern and Geneva, Switzerland;
MOST, Taiwan;
TAEK, Turkey;
STFC, United Kingdom;
DOE and NSF, United States of America.
In addition, individual groups and members have received support from
BCKDF, CANARIE, Compute Canada and CRC, Canada;
COST, ERC, ERDF, Horizon 2020 and Marie Sk{\l}odowska-Curie Actions, European Union;
Investissements d'Avenir Labex, Investissements d'Avenir Idex and ANR, France;
DFG and AvH Foundation, Germany;
Herakleitos, Thales and Aristeia programmes co-financed by EU-ESF and the Greek NSRF, Greece;
BSF-NSF and GIF, Israel;
Norwegian Financial Mechanism 2014-2021, Norway;
NCN and NAWA, Poland;
La Caixa Banking Foundation, CERCA Programme Generalitat de Catalunya and PROMETEO and GenT Programmes Generalitat Valenciana, Spain;
G\"{o}ran Gustafssons Stiftelse, Sweden;
The Royal Society and Leverhulme Trust, United Kingdom.
 
The crucial computing support from all WLCG partners is acknowledged gratefully, in particular from CERN, the ATLAS Tier-1 facilities at TRIUMF (Canada), NDGF (Denmark, Norway, Sweden), CC-IN2P3 (France), KIT/GridKA (Germany), INFN-CNAF (Italy), NL-T1 (Netherlands), PIC (Spain), ASGC (Taiwan), RAL (UK) and BNL (USA), the Tier-2 facilities worldwide and large non-WLCG resource providers. Major contributors of computing resources are listed in Ref.~\cite{ATL-SOFT-PUB-2021-003}.


\clearpage

\printbibliography
\clearpage

\clearpage 
 
\begin{flushleft}
\hypersetup{urlcolor=black}
{\Large The ATLAS Collaboration}

\bigskip

\AtlasOrcid[0000-0002-6665-4934]{G.~Aad}$^\textrm{\scriptsize 98}$,    
\AtlasOrcid[0000-0002-5888-2734]{B.~Abbott}$^\textrm{\scriptsize 123}$,    
\AtlasOrcid[0000-0002-7248-3203]{D.C.~Abbott}$^\textrm{\scriptsize 99}$,    
\AtlasOrcid[0000-0002-2788-3822]{A.~Abed~Abud}$^\textrm{\scriptsize 34}$,    
\AtlasOrcid[0000-0002-1002-1652]{K.~Abeling}$^\textrm{\scriptsize 51}$,    
\AtlasOrcid[0000-0002-2987-4006]{D.K.~Abhayasinghe}$^\textrm{\scriptsize 91}$,    
\AtlasOrcid[0000-0002-8496-9294]{S.H.~Abidi}$^\textrm{\scriptsize 27}$,    
\AtlasOrcid[0000-0002-9987-2292]{A.~Aboulhorma}$^\textrm{\scriptsize 33e}$,    
\AtlasOrcid[0000-0001-5329-6640]{H.~Abramowicz}$^\textrm{\scriptsize 156}$,    
\AtlasOrcid[0000-0002-1599-2896]{H.~Abreu}$^\textrm{\scriptsize 155}$
% ...
\bigskip
\\

% ...
$^{27}$Physics Department, Brookhaven National Laboratory, Upton NY; United States of America.\\
% ...
$^{33}$$^{(a)}$Facult\'e des Sciences Ain Chock, R\'eseau Universitaire de Physique des Hautes Energies - Universit\'e Hassan II, Casablanca;$^{(b)}$Facult\'{e} des Sciences, Universit\'{e} Ibn-Tofail, K\'{e}nitra;$^{(c)}$Facult\'e des Sciences Semlalia, Universit\'e Cadi Ayyad, LPHEA-Marrakech;$^{(d)}$LPMR, Facult\'e des Sciences, Universit\'e Mohamed Premier, Oujda;$^{(e)}$Facult\'e des sciences, Universit\'e Mohammed V, Rabat; Morocco.\\
$^{34}$CERN, Geneva; Switzerland.\\
% ...
$^{51}$II. Physikalisches Institut, Georg-August-Universit\"{a}t G\"ottingen, G\"ottingen; Germany.\\
% ...
$^{91}$Department of Physics, Royal Holloway University of London, Egham; United Kingdom.\\
% ...
$^{98}$CPPM, Aix-Marseille Universit\'e, CNRS/IN2P3, Marseille; France.\\
$^{99}$Department of Physics, University of Massachusetts, Amherst MA; United States of America.\\
% ...
$^{123}$Homer L. Dodge Department of Physics and Astronomy, University of Oklahoma, Norman OK; United States of America.\\
% ...
$^{155}$Department of Physics, Technion, Israel Institute of Technology, Haifa; Israel.\\
$^{156}$Raymond and Beverly Sackler School of Physics and Astronomy, Tel Aviv University, Tel Aviv; Israel.\\
% ...
$^{a}$ Also at Borough of Manhattan Community College, City University of New York, New York NY; United States of America.\\
$^{b}$ Also at Bruno Kessler Foundation, Trento; Italy.\\
$^{c}$ Also at Center for High Energy Physics, Peking University; China.\\
$^{d}$ Also at Centro Studi e Ricerche Enrico Fermi; Italy.\\
$^{e}$ Also at CERN, Geneva; Switzerland.\\
% ...

\end{flushleft}


\end{document}